# Imperial College London

# Tissue characterization based on the analysis on i3dus data for diagnosis support in neurosurgery

## Acknowledgments

I would like to firstly thank my supervisors Dr Stamatia Giannarou (Matina) and Professor Guang-Zhong Yang. Matina has given me a lot of extremely patient guidance in all aspects of doing research, without her I would already end up in some dead end for so many times. I also corrected my bad research routine because of her help, since I used to keep trying new stuff in experiments and completely lost the track and just got frustrated. Professor Guang-Zhong Yang also has given me advices when Matina was away in the beginning and I really appreciate that because I was quite worried for a while.

I would also like to thank our clinical collaborators at Charing Cross Hospital, Mrs. Sophie Camp, Mr. Giulio Anichini and Mr. Dipankar Nandi. They are all occupied by very busy clinical tasks so it's really appreciated that they could help us to provide the ground truth. Especially for Giulio, he was always there whenever I need to ask him clinical questions regarding the project and thanks for the coffee he bought me in meetings.

Thanks to all the colleagues in Hamlyn Centre for all the kind helps I received. Thanks for advices from Xiao-Yun Zhou and Yun Gu regarding deep learning, also thanks so much for Lin Zhang's whole afternoon's effort for installing the system and software on the workstation. And thanks for all my classmates, it's been a great year with you guys.

I greatly appreciate Professor Daniel Elson for writing me a recommendation letter, thanks for organising this MRes course and all his very effective and helpful replies in emails.

I gratefully acknowledge the Hamlyn bursary received from Hamlyn Centre, I would never survive in a city like London without that generous financial support.

# Abbreviations

GB        Glioblastoma

BTR       Brain tumour resection

CAD      Computer-aided-diagnosis

US        Ultrasound

IUS       Intraoperative Ultrasound

I3DUS    Intraoperative three-dimensional Ultrasound

MRI      Magnetic resonance imaging

CT        computerised tomography

SVM     Support vector machines

DL        Deep learning

2D        Two-dimensional

CNN     Convolutional neural network

RNN     Recurrent Neural Network

FCN      Fully convolutional network

FCL      Fully Connected Layer

ReLU    Rectified linear unit

# Content

# Abstract

Brain shift makes the pre-operative MRI navigation highly inaccurate hence the intraoperative modalities are adopted in surgical theatre. Due to the excellent economic and portability merits, the Ultrasound imaging is used at our collaborating hospital, Charing Cross Hospital, Imperial College London, UK. However, it's found that intraoperative diagnosis on Ultrasound images is not straightforward and consistent, even for very experienced clinical experts. Hence, there is a demand to design a Computer-aided-diagnosis system to provide a robust second opinion to help the surgeons. The proposed CAD system based on "Mixed-Attention Res-U-net with asymmetric loss function" achieves the state-of-the-art results comparing to the ground truth by classification at pixel-level directly, it also outperforms all the current main stream pixel-level classification methods (e.g. U-net, FCN) in all the evaluation metrices.

# Chapter 1  Introduction

## 1.1. Clinical background and motivation

There is an imperative desire to develop or improve treatments for brain tumours, since the amount of people who died because of malignant brain tumours has tripped in the last three decades in developed countries [1] although the current project is focusing on the Glioblastoma (GB). The surgical removal of brain tumours is recognized as the most common initial treatment since it substantially improves overall and progression-free survival of patients in brain tumour resection (BTR) surgery [2] [3].

In BTR, the removal of tissues will result in brain shift, consequently, the pre-operative models based navigation will become highly inaccurate and this will lead to insufficient removal while the amount of removal is positively related to survival rate [3] [4]. Therefore, the real-time intraoperative medical images are more reliable for neurosurgeons to visually inspect the tumours margins. However, even the experienced neurosurgeons find the intraoperative diagnosis of Glioblastoma (GB) tumours margins and relevant white matters tracts are challenging, while the accuracy of manual diagnosis is reported as around 75% [2]. To address the issues in intraoperative diagnosis, the researchers have been developing a new technology utilizing the latest advances in computer vision, this technology is termed as computer-aided-diagnosis (CAD). Hence, the aim of the current project is to develop a new CAD system to automatically diagnosis the GB brain tumours.

## 1.2. Why intraoperative ultrasound

Intraoperative Ultrasound (iUS) permits multiple image acquisitions, and minimally augments operative time [5]. It also allows precise visualization of vital structures, progression of tumour resection, and management of immediate complications. It is user friendly, widely available, portable, and a pragmatic, cost effective alternative to intraoperative MRI (iMRI). Furthermore, it does not require any change to theatre set up. By comparison, iMRI despite the superior image quality, is often prohibitively expensive, with limited availability in the UK. It necessitates complex theatre logistics, dedicated equipment and infrastructure, and prolongs operative time. Due to the merits mentioned, the current project is based on the intraoperative ultrasound images.

## 1.3. Challenges

The first challenge is there is no public available database for intraoperative ultrasound images with ground truth, hence, our database is the first of its kind, consequently, our CAD system is also the first CAD system for GB diagnosis based on intraoperative ultrasound images in the world.

The second challenge which is also the most significant one is the intrinsic intra-subject and inter-subject variability in our data. The same type of GB can look very differently in the US images because they could potentially grow at any positions with any shapes, which is also the original motivation for us to develop this CAD system that we aim to enhance the diagnosis consistency significantly. Even for the same patient, due to there is no standard way of data acquisition, neurosurgeons arbitrarily change the ultrasound probe to adjust the best view of tumours, hence, all acquired images are not at the same planes and they are not acquired at the same angles. Therefore, more diversities of images are introduced. Here are some examples:

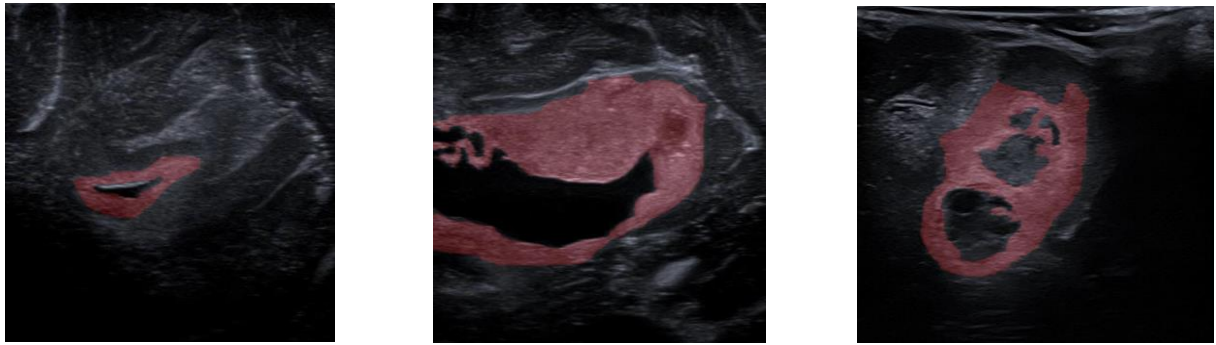

*Figure 1: tumours in intraoperative ultrasound images (red mask: tumour)*

As shown in the above images, the left and the middle ones are from the same patient from the same case and the images were acquired in the same day, the right one is from another case, however, all of them are high grade GB tumours although they look mutually distinctive from each other.

A histogram analysis was also conducted on the acquired images, the programme pairs the US image and the corresponding ground truth from clinical experts, it then sorts the pixels in the US images into either tumour or background classes according to its ground truth. Here are some examples shown in **Figure 2.**

The third challenge is the separation of different classes is not straightforward in our project. First, as shown in above **Figure 1**, the pixels of the two classes are all severely unbalanced across all the acquired images; secondly, the intensities range of the two classes are completely overlapped in all images, which means it's impossible to use thresholding methods, also due to the similarity between the two classes, the CAD system potentially would suffer either extremely high positives or high negatives if the features extractors are not well designed.

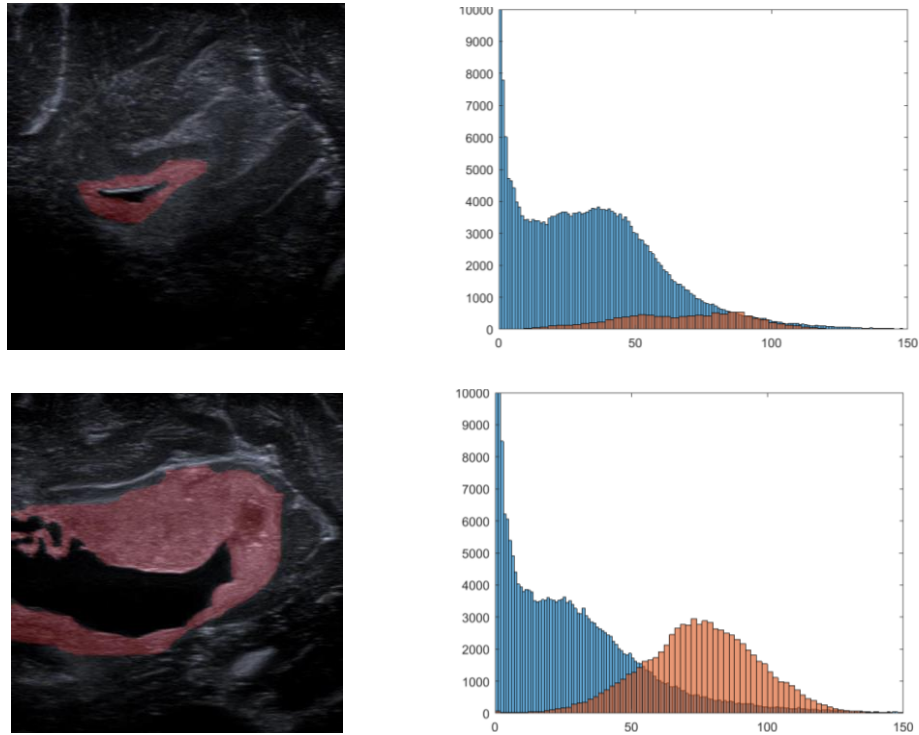

*Figure 2: left above: US of case3 image1; right above: histogram of case3 image1; left down: US of case3 image 9; right down: histogram of case3 image9. In histograms, blue: background; orange: tumours. In US images, red: tumour mask.*

## 1.4. Literature review in computer-aided diagnosis (CAD) system

The CAD emerged in the 1980s at Chicago and has already become one of most popular topics in medical imaging now days [6]. The CAD is defined as a "second opinion" provided by automatic medical image analysis to complement the diagnosis of physicians [6], thus the physicians still need to make the final decisions. Apart from improving the diagnosis accuracy, the CAD also reduces the work burden of radiologists and provides a stable diagnosis performance regardless of inter- and intra-reader variability [1]. The CAD involved with brain tissues is normally designed for two different purposes: detection lesions in brain or binary classification between benign or malignant tumours [1], the later kind is the focus of the present project. In the early days of CAD for BTR, Magnetic Resonance Image (MRI) was the first modality explored, nevertheless, MRI is cost expensive and requires substantial supporting instruments in the operation room [4]. In recent years, the intraoperative ultrasound (iUS) image as an alternative of MRI in CAD has been accepted more and more globally, due to the development of modern US [4]. Unsgaard [4] as well concludes that the iUS empirically achieves same or better image quality comparing to MRI and is suitable for CAD since it can use the same craniotomy from resection. Apart from the inspirations in the literature, the main motivation of the present project stems from a recent observational study undertaken at Charing Cross Hospital, which demonstrated a strong correlation between intraoperative three-dimensional ultrasound (i3DUS) imaging characteristics and histological features. Hence, provided a database of annotated histological i3DUS images, a CAD system to detect brain malignant tumours is highly feasible to achieve.

Generally, a typical CAD system for BTR is composed of two main stages: feature extraction and feature classification, occasionally, pre-processing for data augmentation and post-processing are also involved. The feature extraction stage can be formulated into a pattern recognition problem, therefore modern machine learning techniques have become the centre of the CAD designs, afterwards, the extracted features are fed into classifiers such as SVM, random forest or Softmax.

Before the renaissance of neural networks in 2012/2013, all CAD systems are built on the handcrafted features extractors based on the prior knowledges from the radiologists, the apparent drawbacks include: it's extremely time consuming to design the handcrafted features extractors; those features extractors are very case specialised, not very transferrable to other clinical problems. Due to the extreme excellent performance of neural networks at extracting features, all the very recent CAD systems are using the neural networks, the influence of neural network in medical image analysis is still expanding drastically although the deep learning is already dominating all the medical image analysis methods. Therefore, the literature review is mainly focusing on the deep-learning (DL) based CAD systems.

### 1.4.1. CAD systems based on MRI of brain

Essentially, the final goal of precise diagnosis in medical images is to find the pixel-level label for 2D images or voxel-level label for 3D images. The CNN has started to attract the medical image analysis researchers since it was proposed for natural images classifications, however, due to technical restrictions, most of the frameworks before the appearance of FCN [7] are using a patch based classification which follows a pipeline introduced in last section: patches extraction, train on patches, testing on patches, post-processing. In 2015, a work based on CNN and 3D MRI for brain tumour segmentation was proposed and it introduces a main novelty by using multi-scales inputs. The motivation of that novelty is trying to capture the features as much as possible. Because with patch classification, the label result of the patch is assigned to the centre pixel of the patch, the accuracy of the centre pixel highly depends on the information included in the patch, hence, single sized patches are normally not enough due to the high variability and diversity of brain tumours in images. More specifically, the authors of [8] use inputs including both 2D and 3D at different orientation and different scales in a parallel wise to train a network, the learned features at different scales in different situations are then merged for final representations of features.

Another pioneering CNN based GB brain tumour segmentation framework was proposed in 2016 and the methodology is based on patches-wise classification as well [9]. The main technical novelties include pre-processing and post-processing of MRI images. The pre-processing is scaling the input images by zero-centring and division by standard deviation for normalisation; while the post-processing is thresholding out the small discrete subjects to reduce false positives.

This patches-based classification and its variants are still popular among very recent work [10] [11] [12] [13]. [13] and [10] are other two direct adoptions using patches at single scale and they also successfully segments the brain lesions. The modification in [13] is the network is pre-trained with previously acquired MRI data, essentially, [13] is a domain transferring task; while in [10], the novelty is they use the Res-Net for features extraction. In a landmark based CAD system for Alzheimer's disease diagnosis [12], their network is composed of two stages, the first stage is to locate the most distinctive patches using a data-driven method, the second stage is a typical patches classification, in other words, this method is a patches classification by learning only "useful" patches. Another example of multi-scale inputs for capturing both global and local features is also learning in parallel, however, the authors adopts a features fusion method by directly concatenating the extracted global and local features before the classification layer [11].

Although the patches-based methods are dominating in most of the CAD designs, there are also pioneering works using FCN based method to directly diagnose at pixel-level such as [14] [15].

### 1.4.2. CAD systems based on preoperative US of other organs

The present project is using intraoperative Ultrasound images although there is no CAD system based on intraoperative brain US images for tumour diagnosis in literature, there is only one paper involved with brain [16], however, that work is to segment the brain tissues while our system is for brain tumour diagnosis. Nevertheless, there have been plenty of work using deep learning in US and a very recent literature review in 2018 concludes the statistics of deep learning applications using US in other organs [17]:

| Organ or body location | Modality | Application | # Of papers |
|---|---|---|---|
| Breast | US | Classification (lesions) | 10 |
| Liver | US | Classification (lesions, cirrhosis, other focal and diffuse conditions) | 4 |
| Lung | US | Classification (several diseases, B-lines) | 2 |
| Muscle | US | Classification (atrophy, myositis) | 2 |
| Intravascular | US | Classification (plaque) | 2 |
| Spine | US | Classification (needle placement) | 1 |
| Ovary/uterus | US | Classification (masses) | 2 |
| Kidney | US | Classification (renal disease, cysts) | 1 |
| Spleen | US | Classification (lesion) | 1 |
| Eye | US | Classification (cataracts) | 1 |
| Abdominal cavity | US | Classification (free fluid from trauma) | 1 |
| Thyroid | US | Classification (nodules) | 1 |
| Other | US | Classification (abdominal organs) | 1 |
| Heart | US | Classification (viewing planes, congenital heart disease) Content retrieval | 8 |
| Heart | US | Segmentation | 1 |
| Placenta | US | Segmentation | 1 |
| Lymph node | US | Segmentation | 1 |
| Brain | US | Segmentation | 1 |
| Prostate | US | Segmentation | 2 |
| Carotid artery | US | Segmentation | 1 |
| Breast | US | Segmentation | 1 |
| Gastrocnemius muscle | US | Regression (orientation) | 1 |
| Fetal brain | US | Regression (gestational age) | 1 |
| Spine | US | Registration | 1 |
| Prostate | US (TRUS) | Registration | 1 |
| Liver | Elastography | Classification (fibrosis, chronic liver disease) | 2 |
| Thyroid | Elastography | Classification (nodule) | 1 |
| Breast | Elastography | Classification (tumor) | 3 |
| Liver | CEUS | Classification (lesions) | 1 |
| Total | | | 56 |

*Figure 3: Deep learning application in Ultrasond images* [17]

The US is a real-time modality with excellent portability and economical price comparing to CT and MRI, hence, it's particularly suitable for intraoperative imaging. However, the US is not perfect because the US image qualities are not the most ideal, especially due to the speckle noises and artifacts. Apart from the intrinsic quality limitations, the intraoperative US probe also only acquires images at limited view angle once at a time. What's more, the image analysis of US requires highly experienced experts and the real-time automatic US image analysers are still in the lack. Although there is an existed literature review of machine learning in US [17], we found the review is not very thorough, especially there is nearly no investigations in pixel-level classification whereas that is the latest trend. Hence, we did our independent literature review in classification in US images at both the patch-wise and pixel-wise.

For classifications on patches in US using deep learning, the pipeline is similar to the previous patch-based methods in other modalities as explained in previous paragraphs. A traditional CAD system reported in Science in 2017 is directly classifying the US images at image-wise to diagnose prostate cancer [18], the network is a AlexNet alike architecture, although this work doesn't give a detailed contour of malignant tissues, it still proves the deep learning has the potential for medical use. A system for classifying abdominal US images proposed in 2017 [19] is a typical patch-based example fine tuning a pre-trained VGG16 network. Another typical example is for detecting thyroid nodules [20], the main contribution is the use of a cascade architecture consisting of 2 networks, the first is to find the positive patches and negative patches, the second is to refine the segmentation results. Researchers also try to use multi-streams networks such as the system for cirrhosis diagnosis [21], the system firstly uses sliding-window to slide across the whole image and classify every window to find the contours of liver capsules, based on the located liver capsule, the system does another classification only on patches around the borders for a fine and accurate result. A special case of patch classification is a framework proposed in 2017 for brain regions segmentation [22] in which the authors only use the patch classification in first step for coarse classification, afterwards, they adopt a voting mechanism to precisely localise the contour of tissues.

After the U-net and Fully Convolutional Network were proposed in 2015/2016, researchers have started to predict directly at pixel-level in US, more introductions of networks are presented in later methodology chapter. One example of direct adoption of FCN successfully detects the infant heart in US [23]. In a CAD for detecting infant brain abnormalities proposed in 2018 [24], the main framework is also a typical U-net but with optimized parameters for depth of channel amounts of the network, they decompose 3D volume input into three 2D projections planes, hence, they can process the 2D images at different orientations using 3 normal U-nets respectively, the results show this decomposition of inputs outperforms direct 3D U-net. Another direct application of U-net is for nerve segmentation whereas the version of the U-net is using a Res-Net for encoder [25]. Since US is a real-time modality, the temporal information sometimes can be used to boost the performance. A framework for real-time prostate segmentation in free-hand is also using U-net as a base architecture [26], the main technical contribution is they combine the temporal information by integrating U-net and RNN, an interconnection branch is introduced to take extracted features at different moments in down-sampling stage. [27] is another example combining the RNN and U-net, the different part is they use U-net first to process images from different moments and those initial segmentations become the inputs for RNN for a finer segmentation.

# Chapter 2   Patch-based classification

In the present project, the location of the malignant tissue is unknown, plus the contextual characteristics of the malignant tissue is unknow either. Hence, the present project is developing a system for both the cancerous tissues detection and tissues classification and the first adopted approach is following the classic path by classifying on patches. A few novelties are introduced in the present project from data augmentation to classification on patches:

- A novel multi-scale multi-direction patch extraction for effective data augmentation
- Exploration of the applications of state-of-the-art deep neural networks on our data
- A new parallel classification method based on multi-scale distinct blocks

## 2.1.   Multi-scale multi-direction Patch generation

The ImageNet database has 14 million natural images with ground truth labels, however, it's very difficult to build a medical image database at that scale. The annotations of medical images require not only professional expertise from highly experienced experts but also long devotion of the experts, especially the pixel-level labelling is extremely time consuming. Besides, the subjective bias could be introduced in manual labelling. The general two approaches to compensate the lack of data are: fine-tuning the pre-trained models on ImageNet and data augmentation. Hence, our training patches extraction method must be able to significantly boost the amount of training data.

The quality of patches is as important as the quantity of patches, they must include the patterns in different situations as much as possible. As pointed out in a recent review in 2017 [28], the malignant tissues classification normally requires both the local and global appearance and contextual information, however, all classical networks are trained with inputs at single scale, for analysing at both local and global senses, previous work normally use multi-streams networks which are trained with different sized inputs respectively [28]. Nevertheless, this is apparently computational inefficient. As inspired by work in [29], the regions proposals are focusing on patches along both vertical and horizontal directions with different sizes, the combinations of different shaped anchors can nearly capture any arbitrary shaped objects. The current work extends normal multi-scale patches extraction by cropping patches along different directions, this is to utilise the contextual information more effectively since it's observed in our images that some parts of the cancerous tissues are grown along single direction rather than in a homogenous way. Hence, we propose to extract patches at different scales along different directions, this is not only for data augmentation but also for capturing the features as effectively as possible. The patch extraction idea is illustrated as in below.

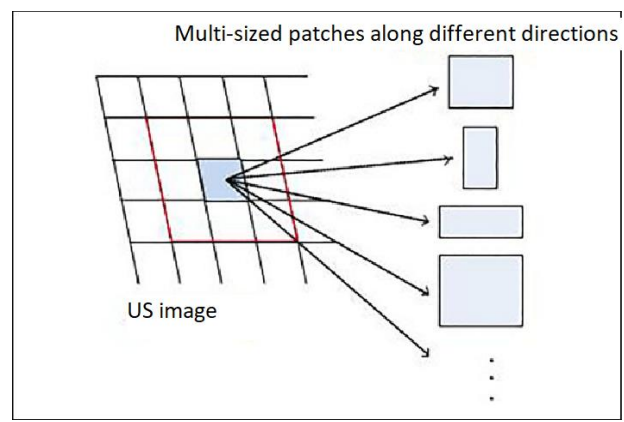

*Figure 4: multi-scale multi-direction patch extraction*

In implementations, a home-made programme is pairing the ground truth binary mask and the corresponding original US first, and then it creates a sliding window for each scale and direction for ground truth. Every sliding window starts from the left-up corner of the ground truth, the sliding step is half the amount of either the width or height of the sliding window, it depends on which direction the sliding-window is going. At each location of the sliding window, if more than 80% areas of the ground truth patch belong to the background, the programme crops the corresponding ultrasound patch at the same location in US and save it as a background patch in background class folder; if more than 80% areas of the ground truth patch are tumour then it will crop the save it into the tumour class folder, all other patches are ignored. This algorithm promises the training patches cropped for each class are not ambiguous. An illustration of locations of saved patches on two US images are also illustrated in **Figure 5** where the red rectangles are for tumour and the blue rectangles are for background.

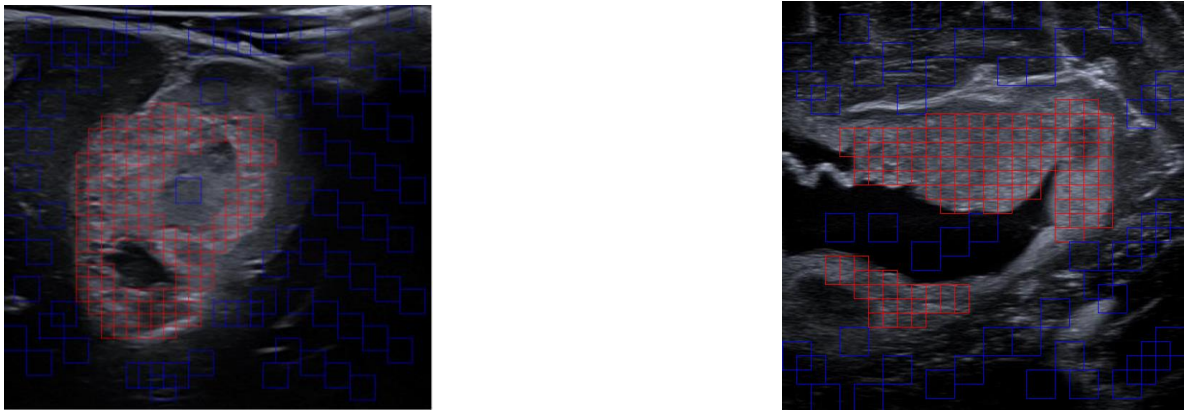

*Figure 5: examples for patches extraction at one scale*

With the method in previous patch extraction, due to the class imbalance, there will be always too many background patches created, for balancing the classes initially, a simple strategy is used to keep only one in every eight background patches and the other background patches are discarded.

## 2.2. Transfer-Learning with pre-trained models

In terms of training methods, a comprehensive analysis in [30] compared three approaches: fine-tuning pre-trained models, training from scratch on medical images and training with handcrafted features. It appears fine tuning method achieves the best results, the main reason might be the amount of the medical images used are way inferior than the requirement of the amount for training all the parameters (e.g. 1.1M parameters in VGG16) in the state-of-the-art neural networks, that's why fine-tuning the pre-trained models always performs better than training models from scratch only on medical images. The current project fine tunes on pre-trained models including: VGG16, GoogleNet (Inception), ResNet-101 and Inception-v2-ResNet.

### 2.2.1. VGG16

The first explored architecture in the current project is the VGG16. The VGG16 shares a similar architecture with previous network AlexNet [31] but with deeper layers, it has 12 convolutional layers and 5 pooling layers, followed by 3 FCLs and a Softmax for classification [32]. The kernels used in VGG16 are smaller than those in previous CNNs (3x3 convolutional layers and 2x2 pooling layers), but larger amounts of channels and more non-linearization mappings, this modification is based on an intuition that two consecutive 3x3 layers can achieve the same receptive filed area of a single 5x5 layer but with more power of representation. Although the VGG16 didn't win the 2013 ImageNet challenge (1$^{st}$ runner-up), it became one of the most popular architectures now days.

### 2.2.2. Inception model

Google research team proposed a module called "inception module" which is in fact a local network topology in 2014, the architecture of the inception module is a combination of clusters of filters bank [33]. Each cluster has either only small sized kernels in parallel or small sized kernels with a pooling kernel in parallel. The ideology of the GoogLeNet is to stack 9 local networks together to make a 22 layers network, however, the authors also suggested to use those modules only in higher layers to improve existed networks (closer to output)[33]. This parallel filters bank design achieves deeper depth without increasing too much computational burden, however, the computational complexity is obviously still an issue [34]. Besides, there is no FCL in the inception modules, for instead, they adopt average pooling to directly generate 1x1x1024 final scores matrix. Another interesting fact about GoogLeNet is they insert two more Softmax layers as "auxiliary classifiers" in intermediate layers to tackle the gradient vanishing problem in back propagation.

### 2.2.3. Residual Net (ResNet)

The most broadly used neural network architecture after 2015 is the ResNet proposed by He et al [35] at Microsoft which also won the 2015 ImageNet challenge. Deeper architecture is not an absolute promise for better performance of a CNN in fact, this is due to the long-existed issue called "vanishing gradients" in neural network, which leads to the worse performances of deeper layers. If we simply stack more layers for identity mapping on a shallow architecture, those extra layers are only copping the shallower layers hence the total training error of the new network will not increase. Normally, the weight layers are essentially fitting the most optimal non-linear relationship ($H$) between the input ($x$) and output ($H(x)$), He et al [35] made a hypothesis that the same layers will find more accurate non-linear relationship ($F$) between the input ($x$) and the residual part of the original output ($H(x) - x$). The proposed residual block has two weight layers for non-linearization ($F$) in cascade with a short-cut connection between the input ($x$) and output ($H(x)$) directly. Essentially, there are two fittings: identity mapping ($h(x)$) and residual mapping ($F(x)$), to achieve the most optimal identity mapping, the weights in the residual mapping will approach all zeros, the precondition is the identity matrix is in the same dimension as the output matrix. With the help of residual blocks, He et al successively pushed the depth of network to 152 layers with lower accumulated training error [35].

## 2.2.4. Inception-ResNet-v2

A combination of Inception modules and Residual modules (blocks) is proposed in 2015 and it achieved the state-of-the-art as well on ImageNet challenge [36]. The introduced difference in the new Inception-ResNet-v2 is there are identity mapping routes added on the original Inception modules, in fact, they did this improvement on their third version of Inception (Inception-v3), with the identity shortcuts, the Inception-ResNet-v2 achieved a much deeper depth.

## 2.2.5. Adoption of pre-trained models

To adopt the pre-trained models on our data, the first modification is replacing all the last fully connected layers in all models to FCLs containing 2 neurons for our binary classification. Due the significant difference between the amounts of parameters in models and the amounts of extracted patches, there is potentially a huge risk of overfitting, to reduce the overfitting, the end structures of the models are changed by adding two more dropout layers with high dropout ratios at 0.7 which is as shown in **Figure 6**.

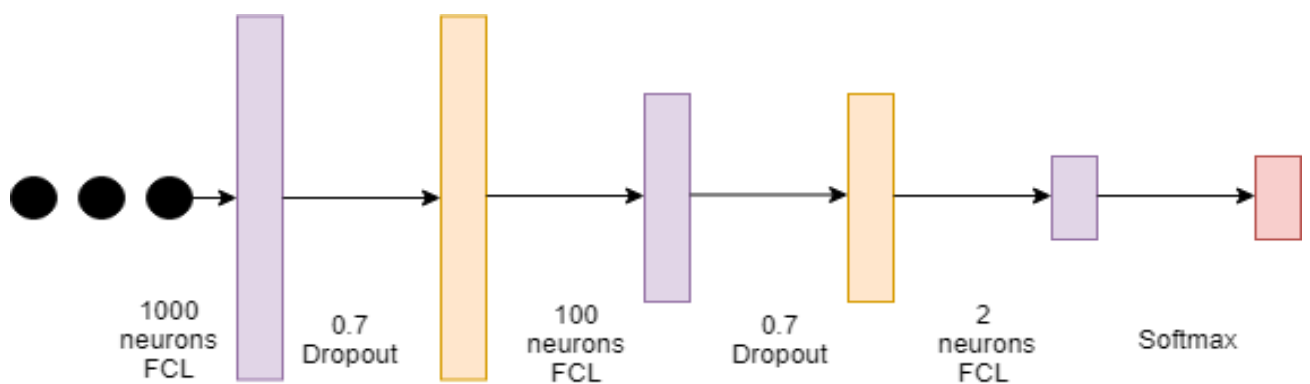

*Figure 6: More dropout layers used in the end of models*

To use the weights more effectively combining with the fact that the lower layers normally extract very similar features regardless of the types of training images, there is no need to fine tune all lower layers of the pre-trained models. In the present project, the VGG16 is relatively shallow so all weights are re-trained using medical image patches; for GoogleNet, all the weights in previous layers are frozen up until the last two inception modules; same for ResNet-101 and Inception-v2-ResNet, all weights are frozen except for the last two composing modules and last fully connected layers.

## 2.3. Multi-scale distinct blocks classifications in parallel

For small tissues classifications, the traditional CAD methods are trying to do pixel-level classification by doing classifications on patches first and then fuse the results together to find the pixel classification at the centre location of the patches, this is effective but with a very expensive computational cost, to speed up classifications with the help of parallel computation, the current project develops a multi-scale distinct blocks classification method. The idea is very straightforward as shown in later **Figure 7**. The testing US image is divided into no-overlapping distinct blocks at different scales, the divided tiles can be classified using the same trained model simultaneously which significantly speeds up the classifications, afterwards, all the classified tiles are reconstructed into the original shape and an average pooling movement is applied for final classification for each pixel of the testing image. The scales used for classifications correspond to scales used for training patches: 20x20, 40x40, 60x60 and 80x80.

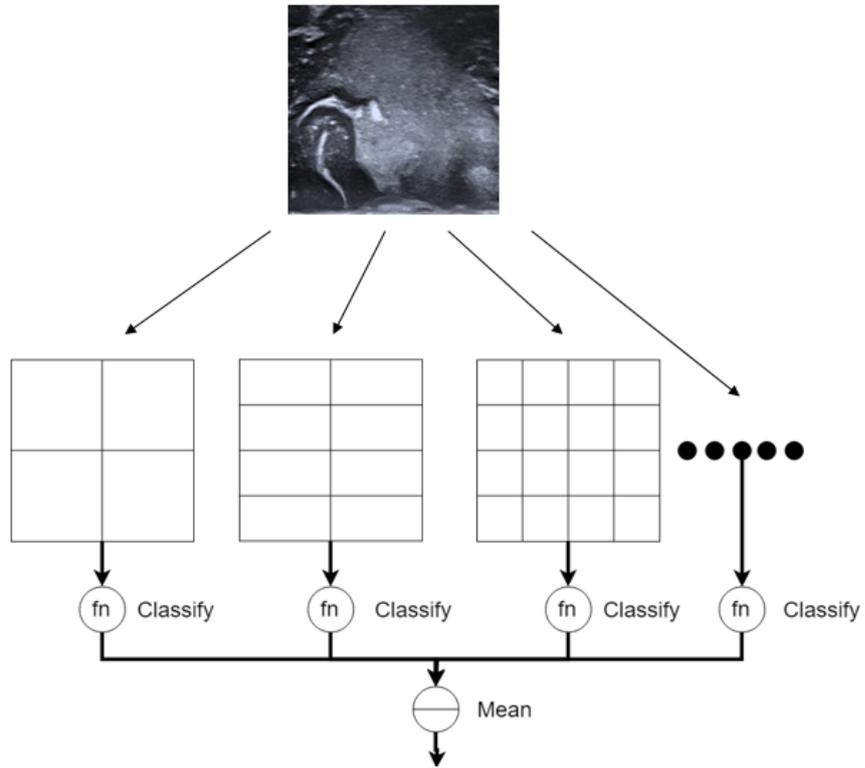

*Figure 7: multi-scale distinct blocks classification with average pooling*

# Chapter 3   Pixel-level classification

The project's scope is focusing on the development of the CAD system to identify the accurate contour of the malignant tissues. To be more precise, it involves with both the locating and classification of the regions of interests. For instead of doing regional classification on divided tiles followed by approximating the pixel labels from the regional classification results, another approach will be doing classification directly at pixel level, in this new perspective, the present project can be formulated into a typical semantic segmentation task in computer vision. Hence, the following chapters focus on how to tackle the problem using the algorithms originally proposed for semantic segmentation.

## 3.1.   Fully Convolutional Network (FCN)

Long et al [37] proposed a framework which excludes fully connected layers for pixel level classification. The FCN is based on the theory: provided the stochastic gradient descent of the whole image is the sum of all gradients of all pixels, it's more efficient to compute the gradients of the whole image than the gradients of patches from one image, the reason behind is due to the high overlapping ratio of receptive fields [37], besides, more than half of the parameters of a network are in fully connected layers, sometimes 90% actually [38]. They argue in their paper that the fully connected layers lose the spatial information of features and force the input size must be fixed [37]. The only difference between a convolutional layer and a fully connected layer is only due to how the hidden neurons connect, in convolutional layer, the neurons are connected to local region of the previous layer while the neurons in fully connected layer are connected to global region of previous layer, however, mathematically, a fully connected layer and a convolutional layer share the same equation to update the weights based on dot products, hence, it's reasonable to assume the fully connected layer can be replaced by normal convolutional layer with careful parameters design as long as the dimensionalities match. As illustrated in [39], if the fully connected layer has 4096 neurons when the input feature map has a size of $7 \times 7 \times 512$ and the output feature map is $1 \times 1 \times 4096$, the fully connected layer can be replaced by a convolutional layer with channel size of $7 \times 7$, padding as 0, stride 1 with 4096 neurons, therefore, the output feature map is kept at $1 \times 1 \times 4096$ after the convolutional layer. By this convolutionalization mechanism, the output will be transformed into a probabilities heatmap including spatial information as shown in **Figure 8**, which naturally forms a semantic segmentation result. In fact, the FCN is more like a module that can be incorporated into existed architectures, for example, as shown in below image, the fully connected layers in a AlexNet alike DAG are convolutionalized, the authors implemented to merge FCN modules into GoogleNet, VGG16 and AlexNet.

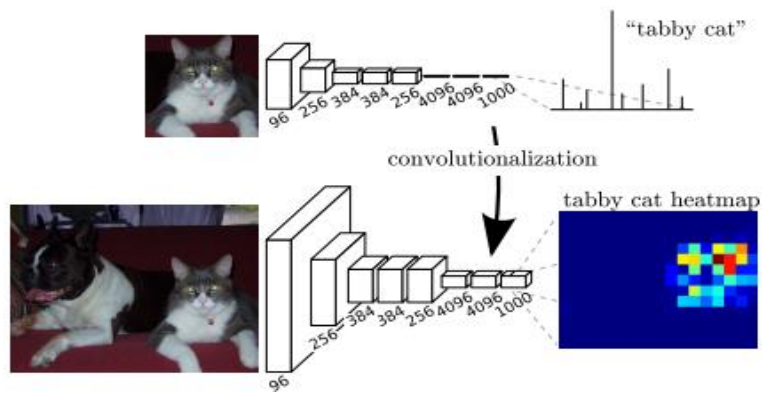

*Figure 8: Convolutionalization of FCL [37]*

Apart from the intuition and the convolutionalization mechanism, another innovation of the FCN work is the combination of spatial information and contextual information. Due to the inherent characteristics of fully connected layer and max pooling layer, these two kinds of layers only propagate the finer and finer contextual information with lower and lower receptive fields to following layers, no spatial information of the features locations are kept. This wouldn't matter in traditional classification problems in image wise, however, with no spatial information, CNN can only do classifications on local regions based on patches, the reconstruction of those fragmented segmentation results rely on a separate method, besides, this kind of segmentation pipeline is extremely time consuming and computational expensive. Long et al [37] came up with the idea to add the coarser feature map to later fine feature map as shown in the figure in below.

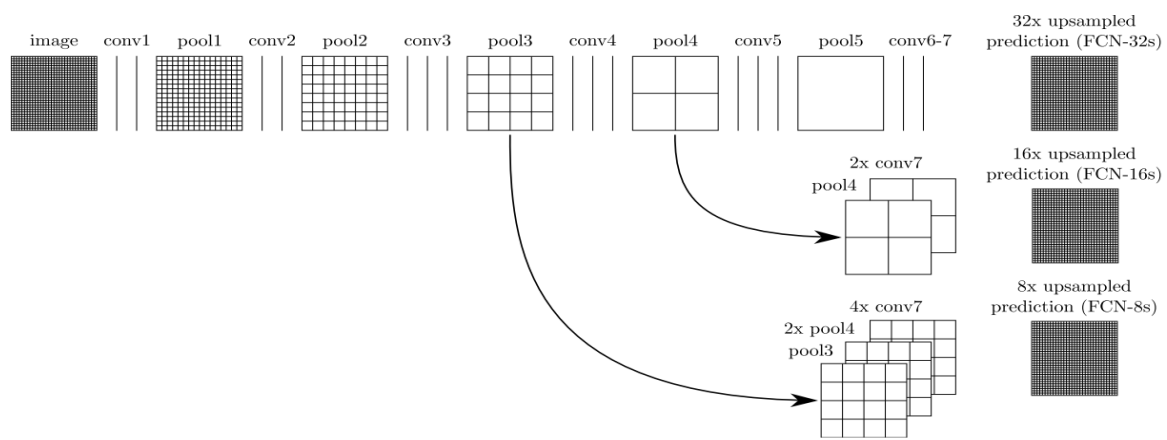

*Figure 9: Combination of coarse and fine features* [37]

The Figure illustrates three methods to combine fine and coarse features at different scales, for example, the FCN-8s is fusing three features maps: 4 times up-sampled pool3 layer, 2 times up-sampled pool4 layer and conv7, those up-samplings are to keep the dimensionalities same as the conv7 for element-wise addition, this setting is also reported as the state-of-the-art in their paper [37].

## 3.2. SegNet

Although FCN inspired a family of neural nets based on the combinations of coarse and fine features using skip-connections since 2015, another work by Badrinarayanan et al [38] gives another direction for semantic segmentation using a encoder-decoder structure inspired by FCN [37] and [40]. In topology, the SegNet has a symmetric architecture as shown in below, it also excludes the fully connected layers like FCN.

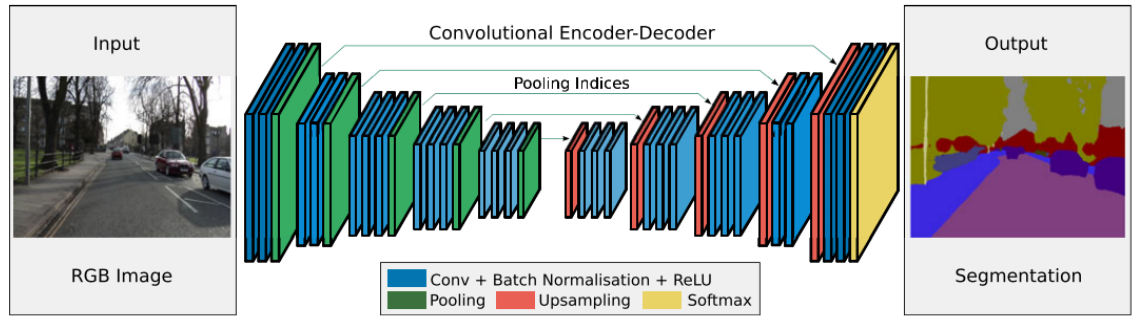

*Figure 10: SegNet architecture* [38]

The encoder part is a pre-trained VGG16 for features extraction in the left, their main contribution lies at the decoder part in which they maintain the indices in max pooling layers in encoder part and guide the un-pooling in decoder part using those corresponding indices, the layers for conducting the guided up-sampling are denoted in red. Each up-sampling layer is followed by three or two blocks consists of convolutional layer, batch normalisation and a ReLU, in the last decoder stage, there is one last Softmax layer for pixel-level classification. The motivation behind is because they also try to recover the lost spatial information and resolution due to the max pooling layers, the difference between SegNet and FCN is, in SegNet, they use an even more computationally efficient way to directly recover the spatial information from the compressed features map rather than appending the spatial information inside of the coarse features on fine features, the whole process is not learnable. The comparison between the indices-guided un-pooling mechanism and the normal de-convolutional mechanism is shown in below.

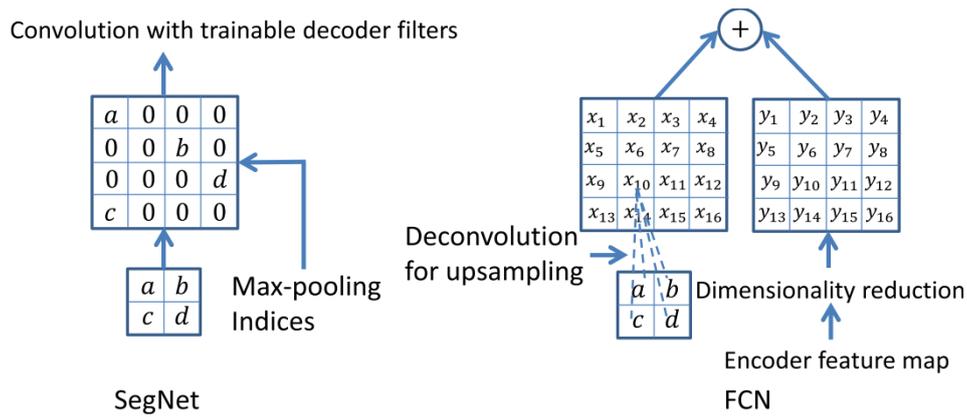

*Figure 11: comparison of indiced un-pooling layer and de-convolutional layer* [38]

As shown in the left-hand side in figure above, comparing to the traditional learnable de-convolutional layers (right hand side), the simply memorizing indices in max pooling are more computationally efficient and straightforward, to compensate the feature capacities in corresponding feature encoder, the following convolutional layers are applied on the up-sampled features, hence the channel amounts are kept same in encoder and decoder. In the SegNet paper, they also tried variant such as the use of "thin" decoder which contains only single channel in decoder comparing to its corresponding encoder stage. Another variant which achieved the state-of-the-art in their experiments is the SegNet-Addition, which is an adoption of "skip connection" idea by adding the up-sampled features and the corresponding features from encoder together. The SegNet-Addition also uses fixed bilinear interpolation for up-sampling, for instance, they tried to up sample all features up to 64 channels at each decoder stage, in the current thesis's author's opinion, this SegNet-Addition is essentially a combination of SegNet and FCN with an increased parameters capacity.

### 3.3. U-net

U-net architecture was proposed in 2015 as a step-change improvement based on FCN and it has been dominating in the medical image community as the benchmark and basement method for segmentation [41]. The original U-net architecture is composed of a down-sampling encoder and an up-sampling decoder as shown in below.

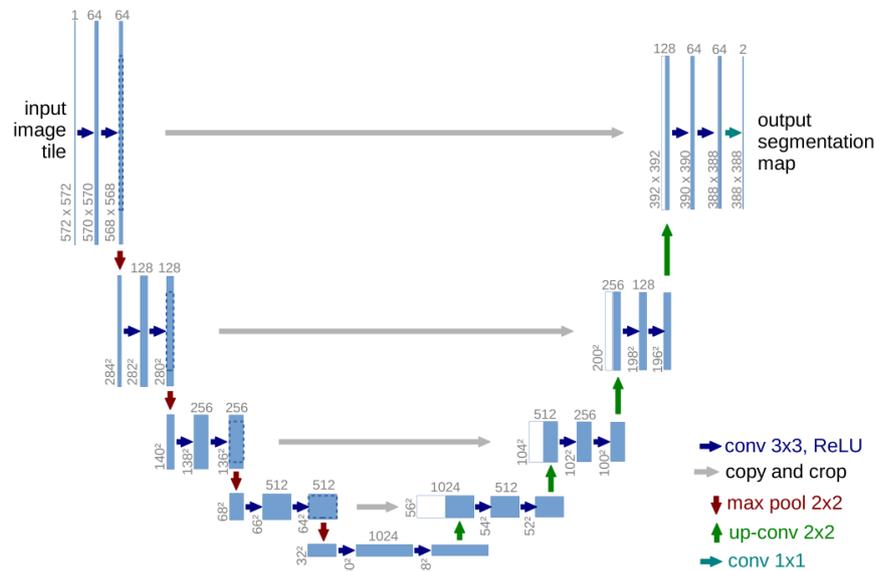

*Figure 12: original U net architecture* [41]

The encoder part [41] on the left hand side is a VGG16 alike cascade of four convolutional blocks (down-sampling stages) which are connected by a stride 2 max pooling layer between each two adjacent down-sampling stages, each block has two 3x3 convolutional layers which are followed by ReLU layers respectively for extracting dense features and non-linearization. Meanwhile, the decoder part has a corresponding cascade structure of four 3x3 convolutional blocks, each block is transited to next up-sampling block (stage) by a 2x2 de-convolutional layer with a factor of 2. The encoder and decoder are connected by a bridge which is made of two 3x3 convolutional layers. No fully-connected layers are used and the U-net can be regarded as one of the FCN, hence, U net can be trained in end-to-end fashion for pixel-level classification as well. As all other fully convolutional networks, U-net can also take arbitrary sized input images as long as the height and width of input images are divisible by the whole down-sampling ratio. For example, in [41], the total down-sampling ratio is $2^4 = 16$ and the height and width of input images are both 572 ($572 = 16 \times 32$).

The main modifications of U-net comparing to FCN is the use of more skip-connections mechanisms at different scales and large number of channels in both down-sampling and up-sampling stages to store more features. The encoder in U net is a typical CNN structure while the receptive filed is expanding with the depth, the locations information of features are lost, plus each max-lpooling reduces the resolution by a factor of 2, after 4 stages of down-sampling, the resolution at the bridge stage is only 32x32 comparing to the input size of 572x572. The authors came up with an idea to compensate the lost locations information and resolution by "copying" and "concatenating" the higher resolution but coarser features from previous stages in encoder to later local finer contextual features in decoder. In other words, the convolutional layers in encoder extract fine local features at different scales, afterwards those local features are merged with global features including higher spatial information at different scales in decoder, unlike the authors of ResNet in which they use element-wise addition to merge global and local features, the authors prefer to "stack" global and local features together to merge them in U-net. The concatenation keeps the global features and local features independently in the same matrix, hence it's more reasonable than simple addition which eliminates the "independency". Another advantage of U-net is due to the use of all pixels in the image hence it requires less training samples, which is extremely important for medical image analysis.

To improve the performance of U-net on our data, the following novel contributions have been introduced and they are explained in detail in later sections:

- Exploration of U-net structures in terms of depth
- Exploration of the use of skip-connections in encoder and decoder

- Comparison of novel Loss functions for classes imbalanced data
- Novel attention mechanism based on human visual system

### 3.3.1. Depth of U-net

Networks with higher capacities perform better at fitting the training data, consequently, higher capacities models come with higher risk for overfitting on validation and testing data. The most primary factor to capacity is the depth of networks. There are no published records of investigations into the depth of U-net. This study aims for finding the most suitable base U-net architecture for the i3DUS segmentation task, which is expected to achieve a conservative performance with the minimum overfitting and the fastest convergence speed.

### 3.3.2. Skip-connections

Since the revolutionary ResNet became the new base CNN architecture for most of recent computer vision tasks from 2015, researchers have been exploring the reason behind the excellence of ResNet. Although the exact theoretical reason is still absent, the whole community is focusing on how to interpret the influences of the skip-connections since that's the main difference between the ResNet and previous state-of-the-art architectures. Drozdzal et al [42] introduced both short skip connections and long skip connections for biomedical image segmentation by incorporating the ResNet and FCN together, they refer all connections in residual blocks as "short" and all connections for recovering spatial resolutions as "long", in other words, "short connections" are for propagating identities of features, while the "long connections" are for propagating coarser features at larger scale to the finer features at smaller scale. Another interesting aspect in their work [42] is they included the identities skip connections in the decoder stages as well, essentially, they used residual blocks in both the encoder and decoder stages, but the result shows a boost of convergence speed in their work.

### 3.3.3. Novel loss functions for classes imbalanced data

This classic cross entropy loss function was originally proposed for natural image analysis. Hence, it has intrinsic drawbacks regarding the medical images. Since in most of circumstances in medical image analysis, the amounts of foreground and background pixels are highly imbalanced. The foreground and background normally get even more severely unbalanced because medical images normally contain much smaller samples in total.

The most straightforward method to address this class imbalance issue is to give different weights to classes, in the present project, the original cross-entropy is used with a weighting mechanism based on the frequencies of each class, where $w$ is the frequency of target class [38], $s_i \epsilon [0\ 1]$ be the score of the $i^{th}$ pixel from the last soft-max layer while $g_i \epsilon \{0,1\}$ is the ground-truth label of the corresponding pixel:

$$Loss = - \sum_{i}^{N} w \cdot g_i \cdot \log(s_i) + (1 - g_i) \cdot \log(1 - s_i)$$

$$w = \frac{amount\ of\ pixels\ in\ foreground}{amount\ of\ pixels\ in\ total}$$

Apart from the weighted cross-entropy loss function as mentioned before, a loss function based on dice overlapping ratio is also applied in [42], this so called dice loss function has been proven to be more robust against the class imbalance in small anatomy segmentation in medical image analysis [43] [44] [45] [46] [47]. The dice coefficient is defined as the similarity measurement between the ground truth (G) and the real segmentation (S) in terms of the mutual overlapping [43] which is also the most common evaluation metric for semantic segmentation:

$$Dice\ score = \frac{2|S \cap G|}{|S| + |G|} = \frac{2\theta_{TP}}{2\theta_{TP} + \theta_{FP} + \theta_{FN}} = \frac{2\theta_{TP}}{2\theta_{TP} + \theta_{AE}}$$

Where the $\theta_{TP}$ is the amount of true positives, $\theta_{FP}$ is the amount of false positives, $\theta_{FN}$ is the amount of false negatives, $\theta_{AE}$ represents all errors. However, this dice score can't be adopted directly since it's not differentiable which is necessary in back-propagation algorithm. A conversion method in [43] shows how to extend the dice score equation to a soft binary loss function which is differentiable for deep learning training. Let's keep using $s_i \in [0\ 1]$ for the score of the $i^{th}$ pixel from the last soft-max layer, and $g_i \in \{0,1\}$ for the ground-truth label of the corresponding pixel, then:

$$\theta_{AE} = \sum_{i}^{N} |s_i - g_i|, \qquad \theta_{TP} = \sum_{i}^{N} g_i (1 - |s_i - g_i|)$$

$$Soft\ Dice = \left(\frac{2 \sum_{i}^{N} g_i \cdot s_i}{\sum_{I}^{N}(s_i + g_i)}\right), \quad Dice\ loss = 1 - \left(\frac{2 \sum_{i}^{N} g_i \cdot s_i}{\sum_{I}^{N}(s_i + g_i)}\right)$$

The back-propagation for the dice loss is:

$$\frac{\partial\ (Dice\ loss)}{\partial\ s_i} = -2 \left(\frac{g_i \sum_{I}^{N}(s_i + g_i) - \sum_{i}^{N} g_i \cdot s_i}{(\sum_{I}^{N}(s_i + g_i))^2}\right)$$

Another version of the dice loss in literature is expressed using a quadratic term [48]:

$$Dice\ loss\ quadratic\ version = 1 - \left(\frac{2 \sum_{i}^{N} g_i \cdot s_i}{\sum_{I}^{N} s_i^{\ 2} + \sum_{I}^{N} g_i^{\ 2}}\right)$$

The corresponding back-propagation for the above equation is [48]:

$$\frac{\partial(Dice\ loss\ quadratic\ version)}{\partial(s_i)} = -2 \left(\frac{g_i (\sum_{I}^{N} s_i^{\ 2} + \sum_{I}^{N} g_i^{\ 2}) - s_i \sum_{I}^{N} g_i \cdot s_i}{(\sum_{I}^{N} s_i^{\ 2} + \sum_{I}^{N} g_i^{\ 2})^2}\right)$$

The above equation is also sometimes expanded into [44] with the addition of $\varepsilon$ term:

$$Binry\ Soft\ Dice\ loss = 1 - \left(\frac{\sum_{i}^{N} g_i \cdot s_i + \varepsilon}{\sum_{i}^{N}(g_i + s_i) + \varepsilon} - \frac{\sum_{i}^{N}(1 - g_i) \cdot (1 - s_i) + \varepsilon}{\sum_{i}^{N}(2 - g_i - s_i) + \varepsilon}\right)$$

Although the benefits of the use of $\varepsilon$ term in [44] are not pointed out, it should be noticed this term is actually a Laplacian smoothing term which is broadly used in Naïve Bayes classification, it's proven to be able to avoid overfitting because the denominator will never fall to zero. This term improves the dice loss function's robustness against the class imbalance. For example, in an extreme situation where all pixels are in background class ($g_i = 0$) and all pixels are classified correctly ($s_i = 0$), without the Laplacian smoothing term, the Binary soft dice loss becomes large which in fact, the loss should be small since all classifications are correct, hence, the gradient will flow to a wrong direction in this extreme situation, however, with the Laplacian smoothing, this "wrong" direction will be kind of fixed.

Since the statistical analysis inspired dice loss function started to get popularity, more investigations were conducted afterwards to find more appropriate loss functions for highly imbalanced biomedical image analysis. Another example is the new loss function called Tversky loss function [45]. The motivation of proposal of Tversky loss function is to furtherly bias the segmentation towards more false negatives for small anatomy comparing to the background, the inspiration comes from the broadly accepted Tversky index for similarity measurement:

$$Tversky\ index = \frac{|S \cap G|}{|S \cap G| + \alpha|S - G| + \beta|G - S|}$$

Where $\alpha$ is for weighting the false positives and $\beta$ is for weighting the false negatives, G is for ground truth (G) and S is for real segmentation.

A more generalised loss function [46] which unifies both the dice loss and Jaccard index was also proposed by the same team proposed Tversky loss function in 2018, this generalised function is based on $F_\beta$ score:

$$F_\beta = (1+\beta^2)\frac{|S \cap G|}{(1+\beta^2)\cdot|S \cap G| + \beta^2|S-G| + |G-S|}$$

The differentiable form is also given along with its back-propagation [46]:

$$F_\beta \text{ loss} = -\frac{(1+\beta^2)\sum_i^N g_i \cdot s_i}{(1+\beta^2)\sum_i^N g_i \cdot s_i + \beta^2 \sum_i^N g_i \cdot (1-s_i) + \sum_i^N (1-g_i)\cdot s_i}$$

$$\frac{\partial(F_\beta \text{ loss})}{\partial s_i} = -\left(\frac{(1+\beta^2)g_i\left(\beta^2 \sum_i^N g_i \cdot (1-s_i) + \sum_i^N (1-g_i)\cdot s_i\right)}{\left((1+\beta^2)\sum_i^N g_i \cdot s_i + \beta^2 \sum_i^N g_i \cdot (1-s_i) + \sum_i^N (1-g_i)\cdot s_i\right)^2}\right)$$

This $F_\beta$ loss function is a uniformed equation for both asymmetric segmentation and symmetric segmentation. When $\beta$ equals to 1, this becomes symmetric dice loss function; when $\beta$ becomes 0, this becomes a Jaccard loss function. Generally, a larger $\beta$ value gives more bias towards false positives and vice versa, this can be played around to find a balance between precision and recall for each special case. Based on the completed soft $F_\beta$ loss function, the current project adopts a mean version of this loss function which is formulated in below.

$$\text{mean } F_\beta \text{ loss} = 1 - \frac{F_\beta \text{ loss numerator} + \varepsilon}{F_\beta \text{ loss denominator} + \varepsilon} \cdot \frac{1}{\text{all observations}}$$

Where $\varepsilon$ is Laplacian smoothing [44].

### 3.4. Attention mechanism

### 3.4.1. Why soft attention mechanism

In most of the computer vision tasks for objects recognitions, the spatial information is extremely important for finding the locations of the objects in the images while all traditional CNN models keep losing spatial information at deep layers. The computer vision researchers have been coming up with different approaches for obtaining the locations of the objects of interests, the most classic algorithms for object recognition are also a family of neural networks based on regional proposal methods such as Fast-R CNN [49], YOLO [50], the main idea of those approaches is they use a bonding box to crop the interested subjects, however, the way to find the location and the size of the bonding box is based on regression rather than learned parameters. All regions based methods can be seen as "hard attention mechanism", the applications of hard attention in medical image segmentation can be grouped into two categories: train an auxiliary branch to learn a regions proposals to locate the objects besides the learning of features in parallel [51] [52]; other approaches are using multi-stages training to progressively reduce the regions of interest [53] [54] [55] [56], they normally start with the whole image, segment and crop the areas including the objects, retrain on the cropped images, repeat this process until the target objects are located. Nevertheless, both these two kinds of approaches are essentially learning more than one network which is not the most efficient method by forcing a "hard" attention.

### 3.4.2. Introduction of attention mechanism for image classification

The soft attention mechanism stems from the new developments in biological neural visual system [57]. In the computational models for focal visual attention [57], a "saliency map" which encodes stimulus conspicuity of the input visual view works very effectively for visual bottom-up "control strategy". In Koch's model [57] which is as show in figure beneath, the information inherited from previous visual processing can be encoded into a "control strategy", afterwards this "control strategy" can guide the system to simplify the current saliency map by reducing the unnecessary saliency, hence, enhance the attention on the most salient location. Their discovery was published before the renaissance of deep learning, therefore, their model used traditional feature extraction (e.g. orientations, intensities, colours et al) methods in parallel for different features in visual processing stage, as illustrated in below, although the feature extractions are top-down flow, the training process is following bottom-up direction.

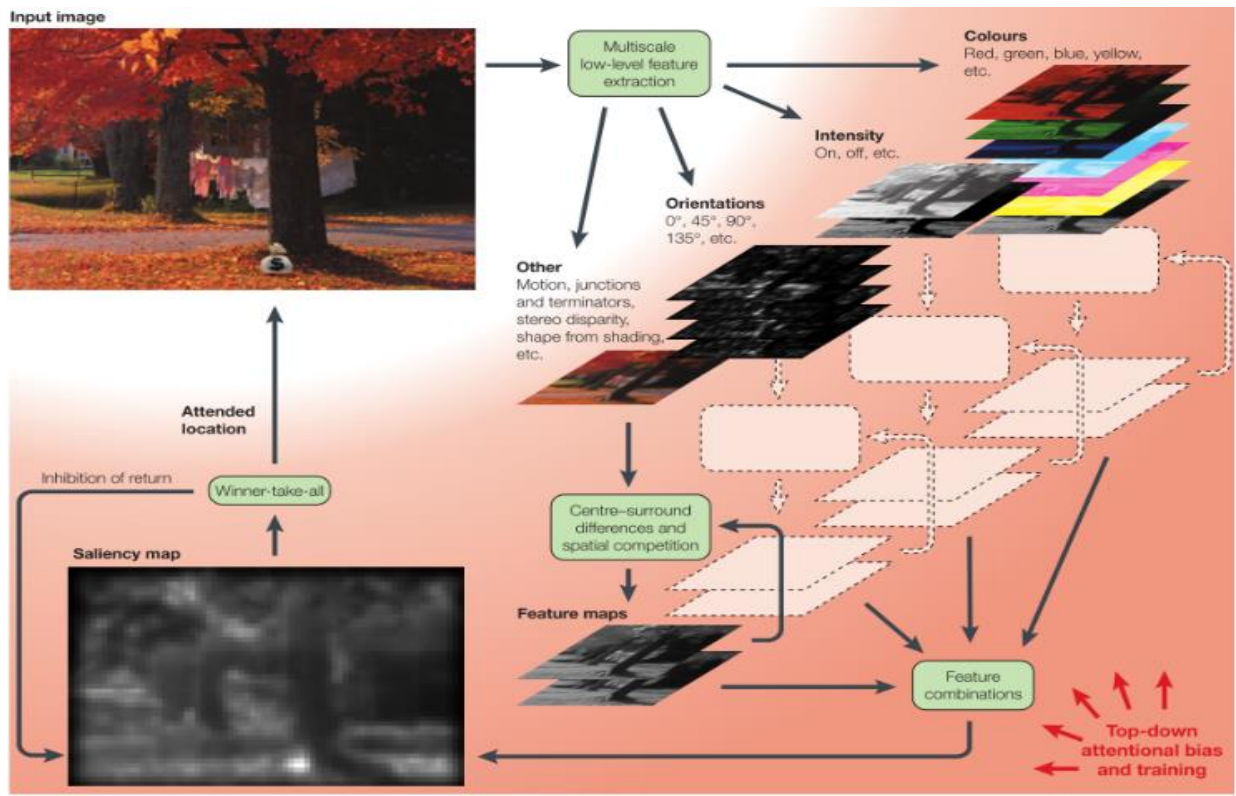

*Figure 13: Computational visual attention model* [57]

A spotlight paper on CVPR 2017 highlighted the effectiveness of the bio-inspired attention mechanism in image classification task which achieved the state-of-the-art performance on ImageNet dataset with 4.8% top-5 error. Want et al [58] successfully adopted the attention mechanism in a feedforward network structure for first time by introducing an attention module as shown in the below.

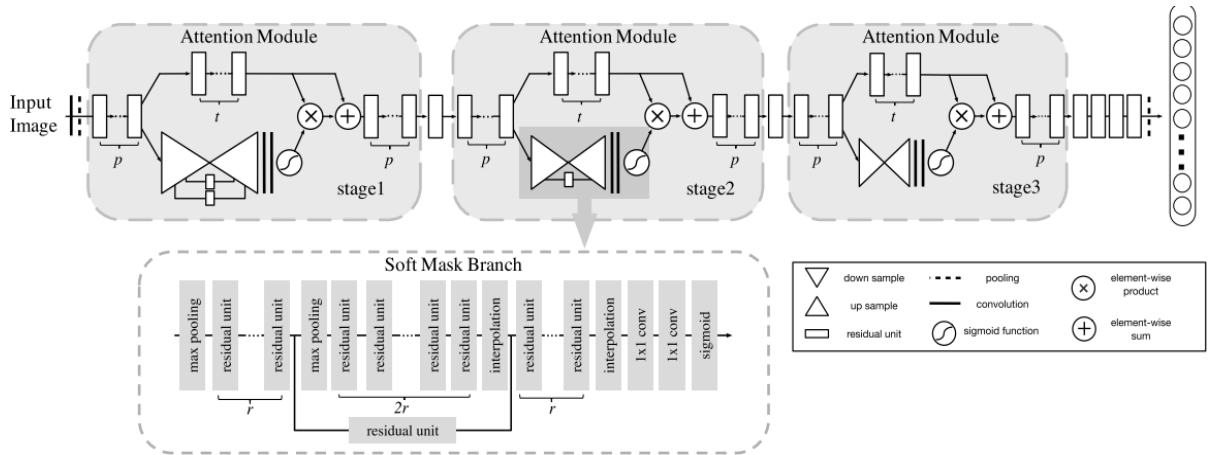

*Figure 14: Residual attention model* [58]

As illustrated in figure above, there are two branches inside of the attention modules, the trunk branch is essentially residual blocks which correspond to visual processing in **Figure 14**, while the soft mask branch outputs the attention weights, it's easily to notice that attention module follows the structure of residual block since the authors want to use the identical mapping idea as well in the attention module. Mathematically, let the input image be $x$, the trunk branch output be $T(x)$ and the mask branch output be $M(x)$, therefore, the attention-aware features as the output of the attention module ($H$) is: $H(x) = (1 + M(x)) * T(x)$. Due to the down-sampling and up-sampling processes inside of the soft mask branch, the effective receptive fields areas in soft branch is also larger than the corresponding trunk branch. They also argue [58] the soft mask branch can not only select features but also suppress the noises from the trunk branch. A comprehensive comparison among different attention mechanisms was also conducted in their work, they compared the attention mechanisms on spatial only, channel only and both the spatial and channel. The different attention mechanisms depend essentially on how to achieve the normalisations as attention weights. In [58], for mixed attention, they simply apply sigmoid function for each channel and spatial locations; for channel attention, they perform L2 normalization for each spatial location across all channels, hence, the attention weights achieved are not related to the spatial information; the spatial attention weights are obtained by applying normalization for each channel followed by sigmoid function.

A more detailed attention mechanism was described in a paper published at ICLR 2018 [59]. The attention module in their work also follows the basic principle of the attention mechanism in [57]: produce the attention weights from the fine global features then add those weights on coarse features to highlight the good saliency. The major difference between the [59] and [58] is the locations of attention modules in the base architectures, besides, the normalisation functions inside of the attention modules are also different. The schematic of their network is shown in below.

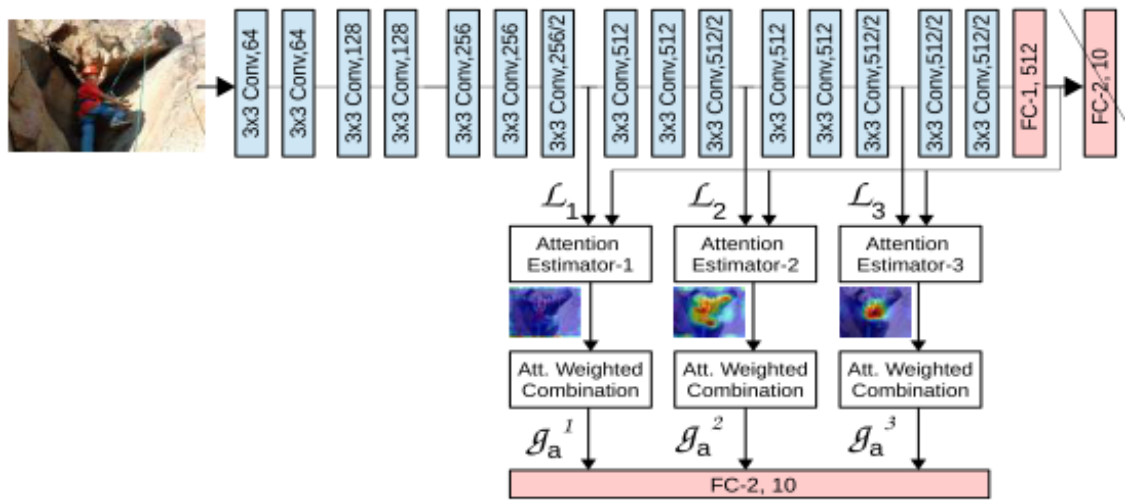

Figure 15: learn to pay attention network [59]

As shown in above, they get rid of the original finally fully connected layer and use the global fine features from the first fully connected layer for calculating the attention weights. Let the coarse features from each layer be $L^s = \{l_1^s, l_2^s, \ldots l_n^s\}$, $s \in \{1, \ldots, S\}$, where s is the index of the layer, $l_n^s$ represents the feature at $n$ pixel location in the layer s, the fine global features is denoted as $g$, however, the dimensionality of the global features is sometimes not the same with the local features, a linear mapping is needed sometimes to match the fine features and coarse features. The attention estimator in figure above is the calculation process to find the compatibility function C, this compatibility function takes the coarse features and fine features as inputs and it outputs the "combination" at a certain dimensionality, it's defined as:

$$\hat{l}_n^s = l_n^s * (linear\ mapping)$$

$$c_i^s = \langle u * \hat{l}_n^s + g \rangle, i \in \{1, \dots, n\}$$

In their work, they simplified the concatenation of $u * l_n^s$ and $g$ to element-wise addition, the weights vector u is learnt to match the dimensionality of fine features and coarse features. The attention weights are essentially the normalisations of the compatibility scores with the use of a Softmax function rather than a sigmoid function as in [58]:

$$a_i^s = \frac{\exp(c_i^s)}{\sum_i^n \exp(c_i^s)}, i \in \{1, \dots, n\}$$

$$g_a^s = \sum_{i=1}^n a_i^s \cdot l_n^s$$

Where the $g_a^s$ is the attention-weighted coarse local features, in their network based on VGG16, all those attention-aware coarse features are then concatenated together to produce the final features. The same attention mechanism was later adopted for detecting ultrasound scan plane and it achieved a substantial improvement [60], hence, it's proven this attention mechanism has a great potential in medical image analysis. Later the same research group extended the use of attention from VGG alike network to U net [61]. This newly proposed attention U net has an architecture shown as:

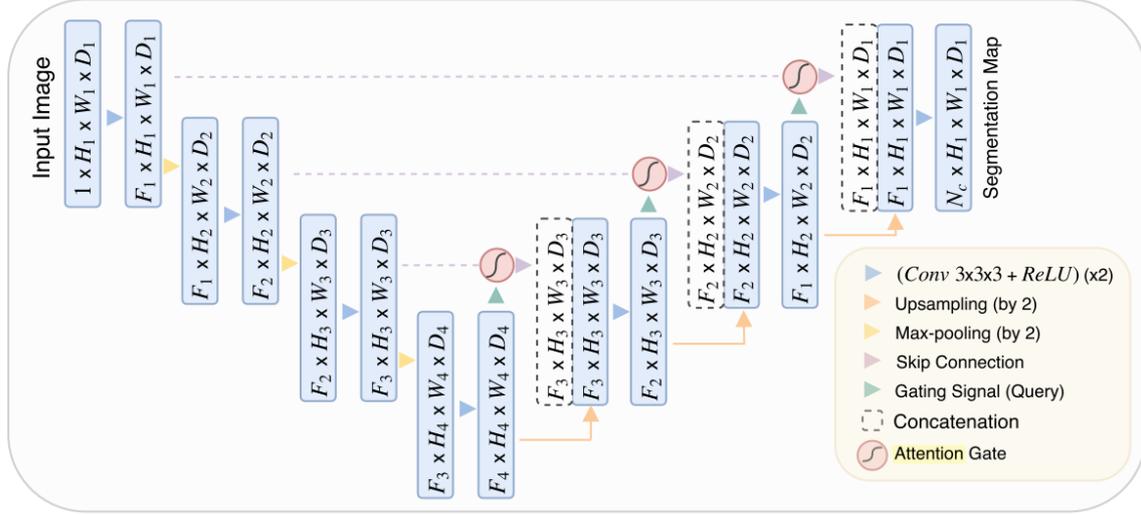

Figure 16: attention U net [62]

As shown in the figure above, the attention gate works the same as the attention module in the [59]: there are two inputs into the attention gate, one is the coarse features from the encoder stage in the long skip connection, the another one is the fine features from the corresponding de-convolutional layer in the decoder stage, afterwards, a slightly different compatibility function is used in [61] [60] which is indicated as:

$$c_i^s = \Psi \sigma_1 (W_f f_i^s + W_g g + b_g) + b_\psi$$

Where $b_\psi \in R, b_g \in R^{Cint}, \sigma_1 = \max(0, x), \Psi \in R^{Cint}$, here are a few bias terms added comparing to the original compatibility map function. The following normalisation step is done in Softmax in [61], but in [60] they apply zero-centring first then divide the score function by its original values.

With the latest advancements of attention mechanism in image classification in literature, the present project introduces a few novel contributions to enhance the performance specifically for intraoperative ultrasound images of brain:

- A novel mixed-attention
- A novel spatial-attention
- A novel integration method of attention in U-net

### 3.4.3. A novel mixed-attention mechanism

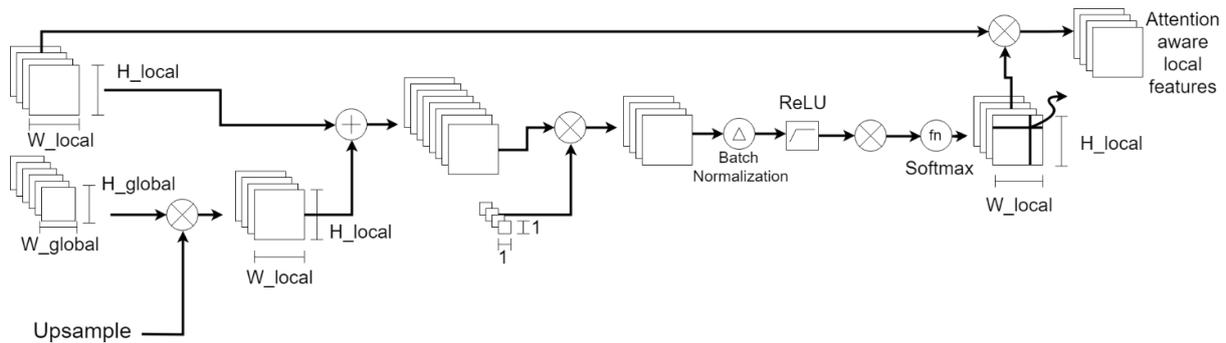

*Figure 17: vector concatenation mixed-attention*

The mixed-attention is simply applying normalisation on all spatial locations for each channel respectively [58], in this sense, our proposed vector concatenation mechanism is also a mixed-attention. As shown in the above figure, first, we project the global features into the local features space by applying a 3x3 transposed convolutional layer on the global features, this transposed convolutional layer doesn't only just up-sample the global features but also build a linear transformation between the global features and the local features, due to this linear mapping, there is no need later to re-sample the attention weights for matching the dimensionality of local features, hence, we don't need to re-sample the attention weights using trilinear interpolation like the previous attention U-net [61], besides, the linear interpolation might bring negative effects to accuracy because it's not learnt during training in [61], plus the spatial distribution of attention weights are not linear either. The transformed global features are then concatenated with the local features, the dimensionality of the features addition is then halved to match with the local features followed by the normalisation.

In terms of normalisation methods for producing attention weights, the previous attentions in U-net [61] and attention in Res-Net [58] are both using sigmoid function while the another attention in a VGG16 alike network is only applying spatial attention using features scaling [60]. However, we propose to use a new normalisation procedure which is different from the previous ones. As it's argued in [60], the Softmax function will potentially increase the sparsity, we propose to scale all the activations at all spatial locations for each channel before the Softmax using zero-centring and division by its own values. In implementation, this normalisation process is done using a batch normalisation layer. After the activations scaling, the ReLU layer plays a gate role to reduce noises, then the scaled and cleaned activations are fed into a Softmax layer to produce the attention weights. As shown in the flow chart in figure, the final dimensionality of the attention weights matrix is the same as the coarse local features input, hence, after the element-wise multiplication between the attention weights and local features, we get the attention-aware local features where the attention weights are applied on all spatial locations for each channel.

This novel attention is termed as "vector concatenation mixed-attention". However, we also compare our attention configuration with two variants:

- Since a previous work implemented the addition using simple element-wise addition [59] for the sake of simplicity, we also compare our vector-concatenation attention mechanism with a corresponding element-wise version, in the implementation of addition attention, we simply use element-wise addition and the 1x1 convolutional layer for dimensionality reduction is also removed after the addition, this variant is referred as "element addition" attention.

- We also compare the current configuration with a variant where the positions of ReLU and the batch normalisation, this move can potentially increase the ability to limit the sparsity because the ReLU filters out the negative half of distribution already before the batch normalisation. This variant is referred as "pre-activation" attention in present project.

### 3.4.4. A novel spatial-attention mechanism

Our second proposed attention is a spatial wise attention mechanism, it shares a similar architecture with the mixed-attention as explained in last paragraph, however, in implementation we switched the positions of the ReLU and the batch normalisation to furtherly prevent the potential sparsity introduced by the Softmax, apart from that, there is one more average pooling on all the channels after the Softmax to acquire only the attentions for each spatial location regardless of the channels.

### 3.4.5. A novel integration method of our attention in U-net

The original motivation for the use of attention is because there are a lot of false positives with the use of original U-net although U-net is the state-of-art architecture for semantic segmentation in biomedical images, then we assume that if we can focus only on the saliency areas of the US, we can significantly reduce the false positives because we simply ignore those low salient areas. Hence, we need to figure a way to integrate the proposed attention mechanism into U-net to enhance its performance.

In the previous attention U-net [61], the coarse local features are from the down-sampling stages in encoder, while the fine global features are from the corresponding up-sampling stages in decoder, hence, the dimensionalities of their two inputs are the same. However, in the interest of the use of the "finest saliency map" for a control strategy of the saliencies in the earlier layers, we should use the features map immediately after the end of the encoder. Because at the end of the encoder, there are the purest and highest dimensional features containing the finest saliencies. Whereas in the decoder, the features are at the end of up-sampling stages and they don't necessarily include all the finest saliencies due to two reasons, the first reason is those features are essentially the addition between the fine global features and the coarse local features, plus the up-sampling process is also a feature selection by reducing the dimensionality, hence, the finest features in the decoder is less than the finest features at the end of the encoder. Therefore, we propose to integrate the attention in the long-skip connections but with the use of features from the end of the encoder for fine features input for all the attention modules, the illustration is shown in below, this is our proposed mixed-attention U-net with short-skip connections in the encoder, which we also refer as "Attention Res-U-net with asymmetric loss function".

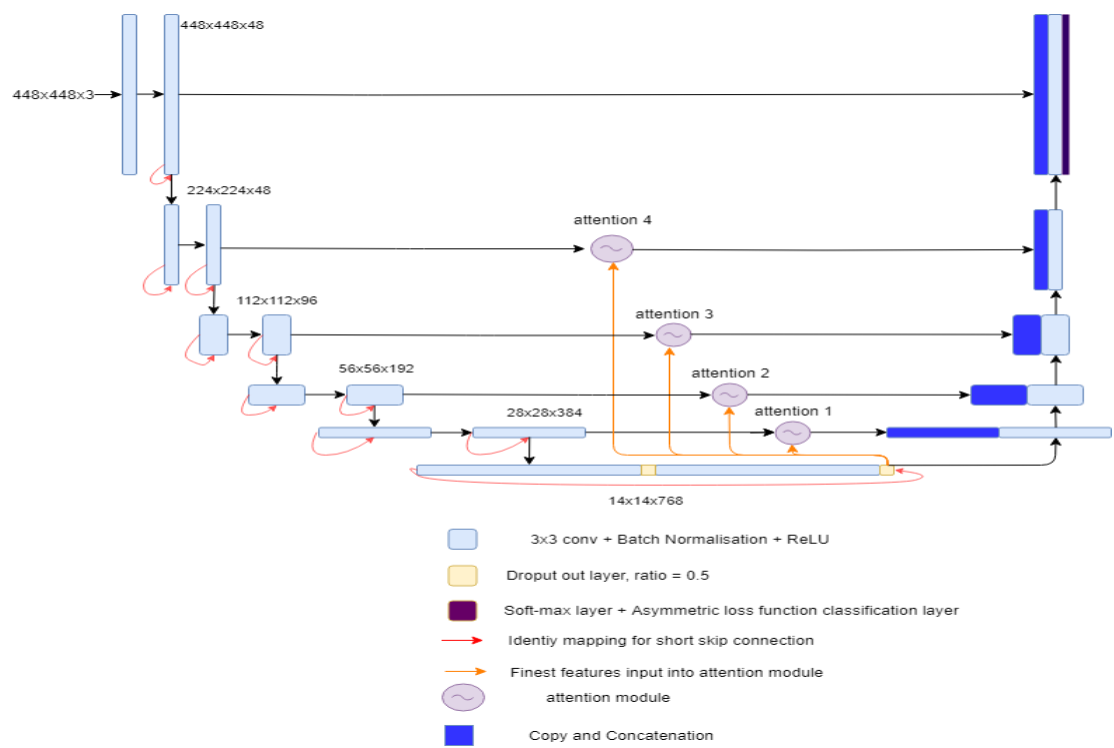

*Figure 18: mixed-attention U-net with short-skip connections in encoder*

# Chapter 4 Experimental results

## 4.1. i3DUS dataset

The initial US video clips were captured using a Toshiba Aplio i800 (Canon Medical System Ltd, Japan) intraoperatively by our clinical collaborators Mrs. Sophie Camp, Mr. Giulio Anichini and Mr. Dipankar Nandi, at Charing Cross Hospital, Imperial College London, UK. The ground truth segmentations were also provided by our same clinical collaborators. Each video lasts around 2 minutes, we exported the videos into image sequences and chose 1 image from every 18 or 20 continuous images, we ended up with 9 or 10 images for each case. Although we have 6 cases up-to-date, we chose to use three cases of the same type of tumour which is GB. Overall, two different high-grade GB cases are used for training and another one GB case is used as unseen case for testing.

For the sliding-window approach, we created 1,322 training small patches for each class using multi-scale multi-direction patch extraction, then we randomly split these patches into two parts: 790 patches per class for training and the rest for validation during training. All patches are resized to 224x224x3.

For the semantic segmentation approach, we created 76 large patches for training, then we randomly split these patches into two parts: 52 patches for training; 24 patches for validation during training. All patches are at 448x448x3 resolution. Although there is a significant reduction in the patch amount, it should be noticed that there are less than 40 training images used in published literature, for example the original U-net and other related work [41] [42]. We are using similar numbers of patch size and patches amount with published paper, hence, the results in the present project are still comparable and meaningful.

The data augmentation is used for all the training including: random scaling along either X or Y axis with factor between [0.3, 4]; rotation randomly between -30 degrees and 30 degrees; random mirroring along either X or Y axis.

## 4.2. Classification

After we get the probability map which is essentially the activation map from the last layer in testing, we adopt this method to assign the labels for pixels in segmentation: Let $s_i \in [0\ 1]$ be the score of the $i^{th}$ pixel from the last soft-max classification layer, $p_i \in \{0, 255\}$ be the intensity of the $i^{th}$ pixel of the final illustration of semantic segmentation, afterwards, this 2-D semantic segmentation is repeated 3 times to form an RGB segmentation map for better visualisation:

$$p_i = \begin{cases} 0 & if\ s_i < 0.5 \\ 255 & if\ s_i \geq 0.5 \end{cases}$$

The sliding-window is using multi-scale distinct blocks which is explained in previous section, segmentation at pixel level is more straightforward. Since there are no fully connected layers used in the architectures in experiments, the semantic segmentation networks can take arbitrary images as long as the sizes parameters are divisible by the down-sampling ratio and the testing image is larger than the training samples. In implementation, the testing images are resized to the size of the training patches which is $448 \times 448$ before they are fed into the trained models, then a simple thresholding is applied on the activations maps from the last layer to get the final semantic segmentation.

## 4.3. Evaluation metrices

Although the classification and training are done at patch level, the final purpose of the system is to detect and classify the tumours tissues, hence, the evaluation of the results are conducted at pixel level. For pixel level evaluation, there are two main branches: direct pixel level accuracy and its variants; intersection over union and its variants.

For the direct pixel level evaluation, the first metric used in the project is simply a ratio between all the correctly classified pixels and the total amount of all the pixels, the present project refers it as global accuracy. It's formulated in equation below.

$$global\ accuracy = \frac{\sum_{i=1}^{m} P_{bb} + \sum_{i=1}^{n} P_{tt}}{\sum_{i=1}^{m} P_{bb} + \sum_{i=1}^{k} P_{tb} + \sum_{i=1}^{w} P_{bt} + \sum_{i=1}^{n} P_{tt}}$$

Where $P_{bb}$ is for background pixel correctly classified as background, true positives; $P_{tt}$ is for tumour pixel correctly classified as tumour, true negatives; $P_{tb}$ is for tumour pixel wrongly classified as background, false negatives; $P_{bt}$ is for background pixel wrongly classified as tumour, false positives.

The second metric is the average of pixel accuracies in each class which is formulated in beneath.

$$mean\ accuracy = \frac{1}{2} \cdot \left( \frac{\sum_{i=1}^{m} P_{bb}}{\sum_{i=1}^{m} P_{bb} + \sum_{i=1}^{w} P_{bt}} + \frac{\sum_{i=1}^{n} P_{tt}}{\sum_{i=1}^{k} P_{tb} + \sum_{i=1}^{n} P_{tt}} \right)$$

As for semantic segmentation, the most frequently used metric is defined as a similarity measurement in terms of the ratio between the intersection and the union of two sets, hence it's referred as Intersection over Union (IoU), sometimes it's also termed as Jaccard Index. Each IoU score is based on a specific class at a time, this present project evaluates the mean IoU of both the classes.

$$Mean\ IoU = \frac{1}{2}\left(\frac{\sum_{i=1}^{m} P_{bb}}{\sum_{i=1}^{m} P_{bb} + \sum_{i=1}^{k} P_{tb} + \sum_{i=1}^{k} P_{bt}} + \frac{\sum_{i=1}^{n} P_{tt}}{\sum_{i=1}^{n} P_{tt} + \sum_{i=1}^{k} P_{tb} + \sum_{i=1}^{k} P_{bt}}\right)$$

A variant of IoU called mean IoU is also used in the present project which is a weighted version of IoU to against imbalanced data [63], where B stands for amount of background pixels, W is the amount of whole pixels, T stands for the amount of tumour pixels.

$$Weighted\ IoU$$

$$= \frac{B}{W} \cdot \frac{\sum_{i=1}^{m} P_{bb}}{\sum_{i=1}^{m} P_{bb} + \sum_{i=1}^{k} P_{tb} + \sum_{i=1}^{k} P_{bt}} + \frac{T}{W} \cdot \frac{\sum_{i=1}^{n} P_{tt}}{\sum_{i=1}^{n} P_{tt} + \sum_{i=1}^{k} P_{tb} + \sum_{i=1}^{k} P_{bt}}$$

The fifth metric used is based on the mean BF score [64], BF stands for Boundary F1 score where F1 score is dice similarity which is monotonic to Jaccard score for each class. This is a measurement of how much the contour of segmentation aligns with the ground truth's contour.

$$BF\ score\ for\ tumour\ class = 2\frac{Precision \cdot Recall}{Precision + Recall}$$

$$Precision = \frac{\sum_{i=1}^{m} P_{tt}}{\sum_{i=1}^{m} P_{tt} + \sum_{i=1}^{k} P_{bt}}$$

$$Recall = \frac{\sum_{i=1}^{m} P_{tt}}{\sum_{i=1}^{m} P_{tt} + \sum_{i=1}^{k} P_{tb}}$$

### 4.4. Experiments set-up

### 4.4.1. Training schemes

Every model was trained at least 3 times for a fair comparison except for 4.4.8.

For sliding-window and transfer-learning approach, we only train the models effectively and stop them when the models start to saturating, after fine tuning on hyperparameters in preliminary experiments, we use initial learning rate at 0.01, learning rate drop factor at 1e-5, learning rate drop period at 100 epochs, total training period of 120 epochs and mini-batch size of 32. The "Adam" optimizer is used with gradient decay factor at 0.9 and squared gradient decay factor at 0.999.

For semantic segmentation training, after fine tuning, we chose to use the same "Adam" optimizer with the same settings, but initial learning rate at 0.1 and total training period of 250 epoch, mini-batch size of 2. No training rate decay scheme is used.

### 4.4.2. Transfer-learning

The hypothesis is the deeper networks (e.g. ResNet, Inception-ResNet-v2) can generalise the training data much better than the shallow ones (e.g. VGG16 and GoogLeNet). The training materials are explained in previous sections. All the models are trained with weighted cross entropy loss function.

### 4.4.3. FCN and SegNet

The original FCN-8s architecture is directly adopted in the project and the weights in pre-trained VGG16 are transferred to the encoder of the FCN-8s. The plain SegNet using a pre-trained VGG16 as encoder is also adopted in the present project. All models are trained with weighted cross-entropy loss.

### 4.4.4. Depth comparison of U-net

Three configurations of U-net with different amounts of down-sampling stages were compared: 3 stages, 4 stages and 5 stages. All the models use weighted cross-entropy loss function. The hypothesis is the deeper U-net is better at extracting features. The width of U-net is chosen to be 48 channels in the first down-sampling stage, because the literatures normally adopt 64 channels, however, to speed the training, we chose to use a slightly lighter network.

### 4.4.5. Where to add attention in long skip connections in Res-U-net

The work in [42] suggests to use short-skip connections (Residual-blocks) in encoder to make the model converges faster, [65] also suggests to directly use a Res-Net as encoder. However, our previous experiments using Res-Net for transfer-learning prove the original ReNet overfits our data easily. Hence, the moderate method is to add short-skip connections in encoder in our base 5 stages U-net which is chosen due to experiments in **4.4.4.**, preliminary experiments did show the models start to converge in very beginning epochs, hence, we chose to adopt the short-skip connections in encoder, we refer this improved 5 stages U-net as "**Res-U-net**".

There are 5 long skip connections in the base 5 stages Res-U-net, we integrated the attention modules in each of the 5 long skip connections respectively, in the experiments, we tested the effect of single attention module in different long skip connections, in all configurations, the attention modules take the features from the end of the encoder for fine-features input.

### 4.4.6. How much attention should be added

In the previous section, there are only single attention modules added in long-skip connections respectively, on the contrary, we conducted experiments on accumulation effect of multiple attention modules in long skip-connections. We started with only two attention modules in the bridge stage and the first decoder stage. Then we added another one more attention module in the next decoder stage, until the bridge and the next 4 decoder stages are fully integrated with attention modules. In total, we tested the Res-U-net with 2, 3, 4, 5 attention modules respectively. All trained with asymmetric loss function with beta value at 1.

### 4.4.7. Short-skip connections in encoder

To investigate the effect of short-skip connections in encoder in U-net, we compared attention U-net and its corresponding residual version with short-skip connections in encoder. For a fair comparison, two attention U-net configurations were examined, one is the attention U-net proposed in literature [61], another one is attention U-net using end of encoder for fine features inputs. All trained with asymmetric loss function with beta value at 1.

### 4.4.8. Short-skip connections in both encoder and decoder

There is lack of investigations of short skip-connections in the decoder, hence, we did a few investigations in the effect of short skip-connections in the decoder. In this section, we tested the Res-U-net with 4 mixed-attention modules at different beta value in the asymmetric loss function, there are two versions for each configuration: with short skip-connections in decoder and without short skip-connections in decoder. Hence, the only comparable variant here is whether the auxiliary skip-connections exist in the decoder or not. All trained with asymmetric loss function with beta value at 1.

### 4.4.9. Encoder for fine features inputs or decoder for fine features inputs in attention modules

In the previous attention U-net [61], the attention modules is in a grid manner that the fine features inputs are from different scales in the decoder while we argue it would probably be better to use fixed global feature from the end of the encoder, in later sections, the "grid attention" is referring the attention U-net in literature. Two experiments were conducted using two different configurations: Res-U-net and plain U-net, in each configuration, there are two versions: attention modules take decoder features as fine features (proposed in literature); attention modules take end of encoder features as fine features.

### 4.4.10. Comparison between element-addition and vector-concatenation in attention

We compared the element-addition attention modules and vector-concatenation attention modules in plain 5 stages U-net and Res-U-net. To keep the consistency, they both used asymmetric loss function with beta value at 1.

### 4.4.11. Comparison between cross-entropy loss and asymmetric similarity loss (beta = 1)

The cross-entropy loss is not robust against the classes imbalance hence we compared the original cross-entropy and the asymmetric loss function [46]. However, for simplicity, the beta value was kept at 1 across all the experiments in this section, essentially, we were comparing the cross-entropy and the Dice loss function which is a special form of the asymmetric similarity loss function. Two configurations were adopted for this comparison, one is plain U-net with element-addition attentions and the another is plain U-net with vector-concatenation attentions.

### 4.4.12. Comparison between spatial attention and mixed attention

To compare two common types of attention modules, we evaluated two configurations: Res-U-net with 3 attention modules and Res-U-net with 4 attention modules. In each configuration, we tested spatial-attention version and mixed-attention version.

### 4.4.13. Influence of hyperparameter beta in the asymmetric loss function

We also compared different hyperparameter values of the beta in the asymmetric loss function to evaluate how much it controls the false positive and false negatives. We evaluated parameters: 0, 0.5, 1, 1.25, 1.5 and 2, the testing models are Res-U-net with 3 mixed attention modules and Res-U-net with 4 mixed attention modules.

### 4.4.14. Comparison between pre-activation mixed attention and after-activation mixed attention

In side of the attention module, what we proposed is very similar to normal convolutional layers, however, it's known that the order of batch normalisation and non-linear function influences the flow of gradients. Hence, we also compared two different mixed-attention modules: ReLU before the batch normalisation and ReLU after the batch normalisation. The two tested configurations are Res-U-net with 3 attention modules and Res-U-net with 4 attention modules.

## 4.5. Results and Discussion

In this section, the validation accuracy of the tested model is shown in figures and the testing accuracy on unseen case are shown in tables, in the tables, the first value is the mean score and the value in the bracket is the standard deviation.

### 4.5.1. Fine-tune pre-trained models on multi-sized patches

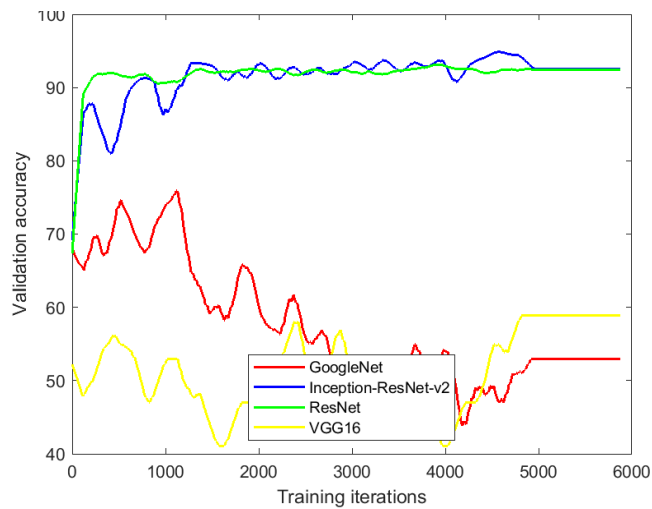

*Figure 19: Fine tuning on pre-trained models, all trained with cross-entropy*

*Table 1: Testing scores on unseen case using transfer-learing and patches*

| Model_names | Global_accuracy | Mean_accuracy | Mean_IoU | Weighted_IoU | Mean_BF_Score |
|---|---|---|---|---|---|
| Inception-ResNet-v2 | 0.56967 (0.019578) | 0.58906 (0.019981) | 0.381 (0.018656) | 0.37383 (0.018651) | 0.19718 (0.029926) |
| Res-Net | 0.60256 (0.010126) | 0.62359 (0.010729) | 0.41056 (0.009097) | 0.40294 (0.0089258) | 0.24311 (0.013963) |

Although the VGG16 and GoogleNet were also fine-tuned, however, in the testing, these two models generate extremely biased results that they recognise all the pixels as either tumour and background, hence, their results are not included due to lack of comparability.

Combining with validation accuracy, the fine-tuned models achieved high validation accuracy more than 90% during training, however, this is probably due to overfitting since the testing accuracy on unseen case is not good.

### 4.5.2. FCN and SegNet

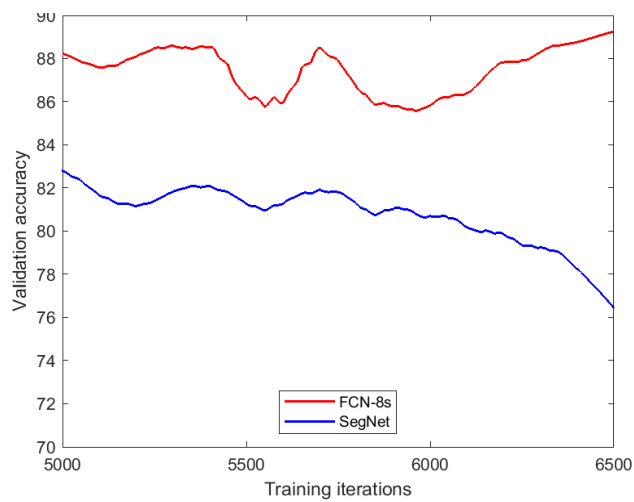

*Figure 20: FCN-8s and SegNet, all trained with weighted cross entropy*

*Table 2: Testing scores on unsee case using FCN and SegNet*

| Model_names | Global_accuracy | Mean_accuracy | Mean_IoU | Weighted_IoU | Mean_BF_Score |
|---|---|---|---|---|---|
| FCN-8s | 0.83817 (0.043294) | 0.83928 (0.048786) | 0.72162 (0.065312) | 0.72247 (0.06379) | 0.3645 (0.101) |
| SegNet | 0.72846 (0.071153) | 0.71737 (0.074913) | 0.56101 (0.10085) | 0.56616 (0.098525) | 0.27335 (0.056009) |

The FCN-8s achieve both higher validation accuracy during training and higher testing accuracy than the SegNet. The standard deviations of the first four metrices of FCN are also smaller than SegNet so the FCN is not only more accurate but also more stable.

### 4.5.3. Depth of U-net

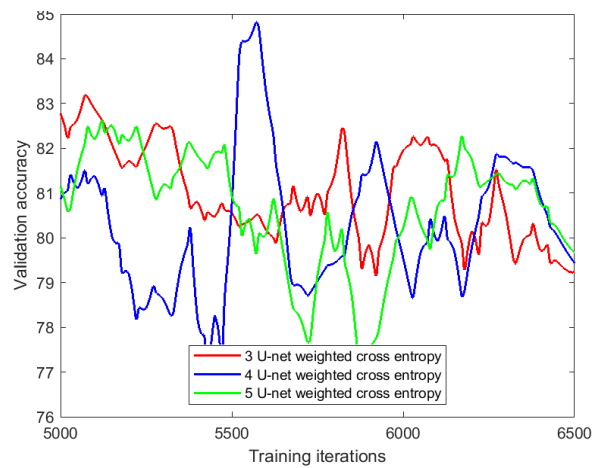

*Figure 21: U-net at different depths, all trained with weighted cross-entropy*

*Table 3:Testing scores on unsee case using U-net at different depths*

| Model_names | Global_accuracy | Mean_accuracy | Mean_IoU | Weighted_IoU | Mean_BF_Score |
|---|---|---|---|---|---|
| 3 stages U-net_weighted_crossentropy | 0.69217 (0.019078) | 0.67109 (0.021383) | 0.49247 (0.029997) | 0.50192 (0.028741) | 0.31473 (0.010962) |
| 4 stages U-net_weighted_crossentropy | 0.68885 (0.07868) | 0.66941 (0.088852) | 0.49151 (0.12154) | 0.50068 (0.11631) | 0.28963 (0.048187) |
| 5 stages U-net_weighted_crossentropy | 0.68145 (0.043264) | 0.65928 (0.04727) | 0.47647 (0.064082) | 0.48659 (0.061625) | 0.30595 (0.032817) |

In the validation accuracy comparison, the 3 U-nets perform very similar while the 4 stages U-net fluctuates more than the other two. In the testing, they still perform very similar to each other while the 3 stages U-net slightly outperforms the other two. However, we still use the 5 stages U-net because the 5 stages U-net extract the abstract features better due to the deeper depth.

### 4.5.4. Single attention module in different long skip connections

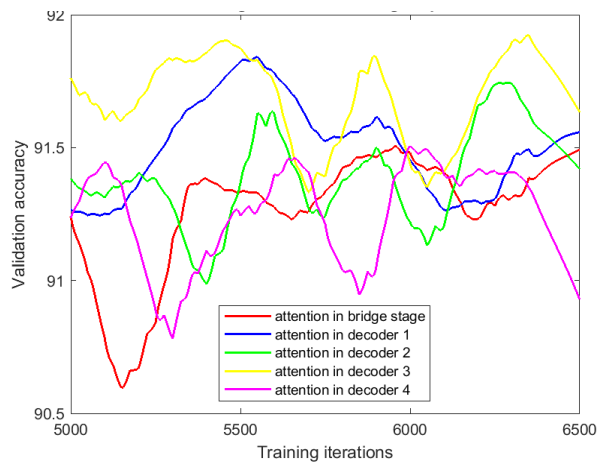

*Figure 22: Effect of single attention modules at different locations, all trained with asymmetric loss function, beta = 1*

*Table 4:Testing scores on unsee case using U-net with single attention*

| Model_names | Global_accuracy | Mean_accuracy | Mean_IoU | Weighted_IoU | Mean_BF_Score |
|---|---|---|---|---|---|
| Res-U-net, attention in bridge | 0.85748 (0.024402) | 0.8653 (0.022959) | 0.75038 (0.037552) | 0.74955 (0.038018) | 0.42558 (0.043515) |
| Res-U-net, attention in decoder 1 | 0.82251 (0.045349) | 0.83293 (0.041484) | 0.69765 (0.069126) | 0.69586 (0.07037) | 0.40511 (0.054137) |
| Res-U-net, attention in decoder 2 | 0.85372 (0.013419) | 0.86139 (0.012549) | 0.74436 (0.020617) | 0.74356 (0.020889) | 0.43071 (0.018886) |
| Res-U-net, attention in decoder 3 | 0.81412 (0.012526) | 0.82582 (0.011768) | 0.68377 (0.018498) | 0.68169 (0.018815) | 0.39512 (0.0076134) |
| Res-U-net, attention in decoder 4 | 0.87229 (0.015346) | 0.87724 (0.013116) | 0.77339 (0.024574) | 0.77321 (0.025118) | 0.43154 (0.038793) |

The performances of the 5 configurations with single attention modules at different locations are very similar to each other in terms of validation accuracy, it seems the attention in the decoder stage 3 brings the most improvement into the network. However, the testing results show different results when it comes to unseen case that the attention module in the decoder stage 4 is the best one, the attention in the bridge stage also performs slightly better than others.

### 4.5.5. Multi attention modules in Res-U-net

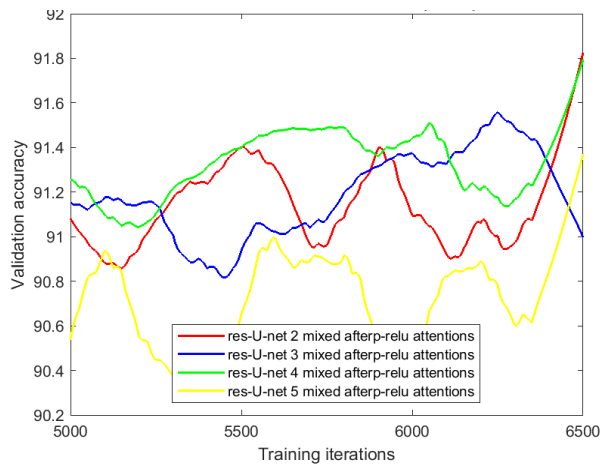

Figure 23: Effect of multiple attention modules, all trained with asymmetric loss function, beta = 1

*Table 5: Testing scores on unsee case using multiple attention modules*

| Model_names | Global_accuracy | Mean_accuracy | Mean_IoU | Weighted_IoU | Mean_BF_Score |
|---|---|---|---|---|---|
| Res-U-net, 2 attention modules | 0.84153 (0.030603) | 0.85058 (0.028567) | 0.72593 (0.04635) | 0.7247 (0.047066) | 0.42498 (0.040125) |
| Res-U-net, 3 attention modules | 0.82784 (0.012076) | 0.836 (0.012276) | 0.70559 (0.017487) | 0.70454 (0.017502) | 0.39025 (0.028309) |
| Res-U-net, 4 attention modules | 0.86548 (0.021367) | 0.87167 (0.019422) | 0.763 (0.033812) | 0.76254 (0.03426) | 0.43682 (0.03707) |
| Res-U-net, 5 attention modules | 0.84031 (0.034766) | 0.84972 (0.032005) | 0.72402 (0.052832) | 0.72267 (0.053783) | 0.40738 (0.047106) |

The validation accuracy differences among the Res-U-net with different numbers of attention modules are not clear although it seems the Res-U-net with 3 attention and the one with 4 attention slightly perform better. However, in the testing on the unseen case, it's more clearly that the Res-U-net with 4 attention modules performs better than other configurations in all evaluation metrices. Hence, we chose to use this Res-U-net with 4 attention modules for our final base configuration and for simplicity, this configuration is referred as "Res-U-net".

## 4.5.6. Effect of short-skip connections in encoder

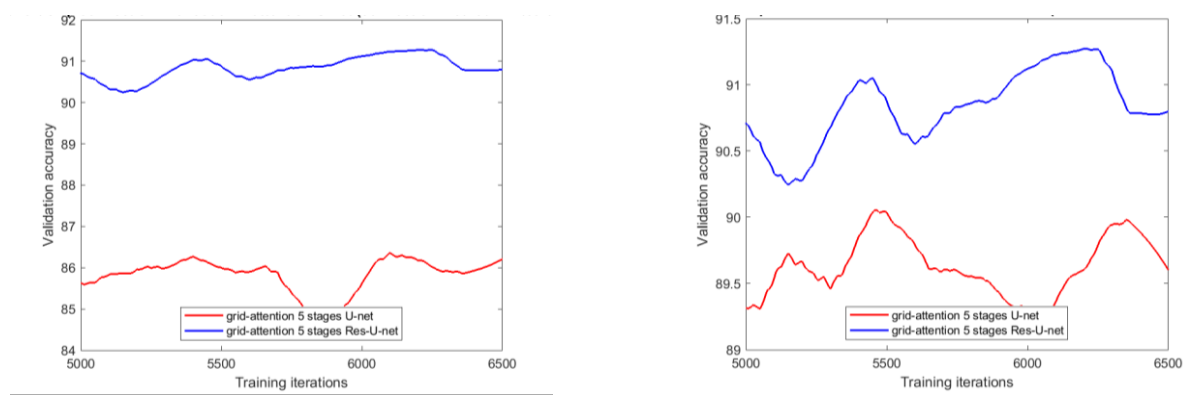

*Figure 24: Left: Attention modules use features from decoder as fine-features inputs; Right: Attention modules use features from the end of encoder as fine-features inputs. All trained with asymmetric loss function, beta = 1.*

*Table 6: Comparison of testing scores on unseen case between U-net with short skip-connections (Res-U-net) in encoder and plain U-net*

| Model_names | Global_accuracy | Mean_accuracy | Mean_IoU | Weighted_IoU | Mean_BF_Score |
| --- | --- | --- | --- | --- | --- |
| U net 4 attention, beta = 1 (connection in literature) | 0.69517 (0.18311) | 0.70369 (0.16239) | 0.52924 (0.23524) | 0.5257 (0.2449) | 0.32201 (0.029123) |
| Res U net 4 attention, beta = 1 (connection in literature) | 0.48723 (0.041398) | 0.51884 (0.036874) | 0.26432 (0.05546) | 0.25009 (0.058381) | 0.40885 (0.24161) |
| U net 4 attention, beta = 1 (proposed connection method) | 0.73711 (0.11586) | 0.74894 (0.10495) | 0.58074 (0.15471) | 0.57736 (0.15981) | 0.33891 (0.082998) |
| Res U net 4 attention, beta = 1 (proposed connection method) | 0.86548 (0.021367) | 0.87167 (0.019422) | 0.763 (0.033812) | 0.76254 (0.03426) | 0.43682 (0.03707) |

The introduction of short-skip connections improves the validation accuracy in both the two tested configurations, whereas there is contradictory result in the testing, with the connection method of attention in literature, the short skip connection decreases the testing accuracy while it improves the accuracy in the models using proposed connection method. Due to the Res-U-net with 4 attention modules is the best performed model, we kept to use this as our base architecture.

## 4.5.7. Effect of Short skip-connections in both the encoder and decoder

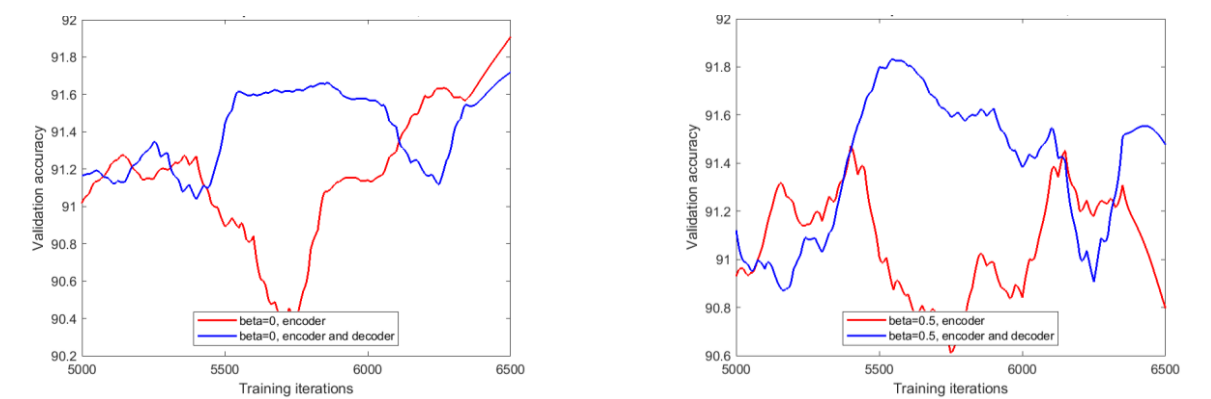

*Figure 25: Left: skip-connections in encoder and skip-connections in encoder and decoder, trained with asymmetric loss, beta = 0; Right: skip-connections in encoder and skip connections in both encoder and decoder, asymmetric loss function, beta = 0.5*

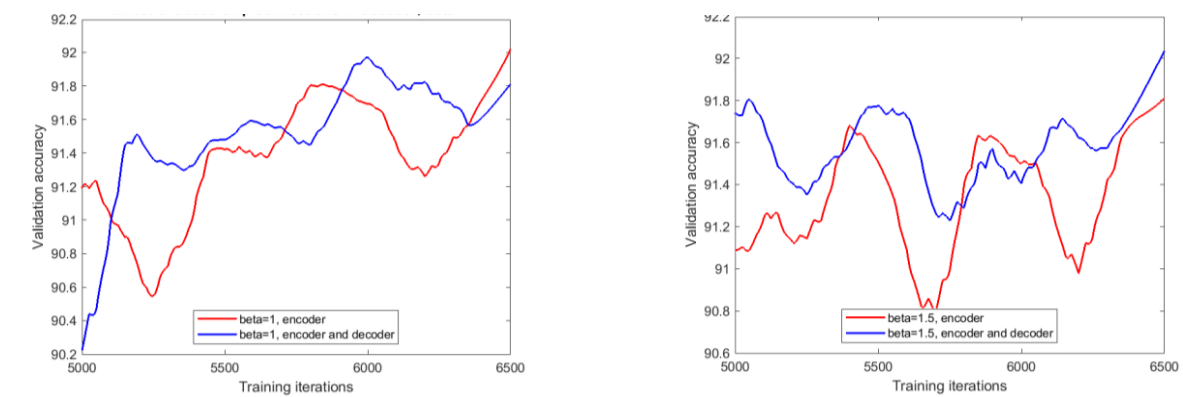

*Figure 26: Left: skip-connections in encoder and skip-connections in encoder and decoder, trained with asymmetric loss, beta = 1; Right: skip-connections in encoder and skip connections in both encoder and decoder, asymmetric loss function, beta = 1.5*

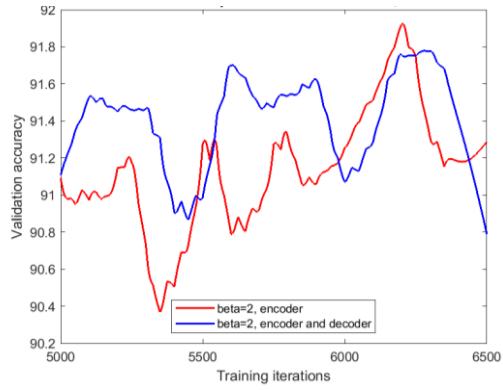

*Figure 27: skip-connections in encoder and skip connections in both encoder and decoder, asymmetric loss function, beta = 2*

*Table 7: Comparison of testing scores on unseen case between U-net with short skip-connections in encoder and U-net with short skip-connections in both the decoder and encoder*

| Model_names | Global_accuracy | Mean_accuracy | Mean_IoU | Weighted_IoU | Mean_BF_Score |
|---|---|---|---|---|---|
| Res U net 4 attention, beta = 0 | 0.8432 | 0.8526 | 0.7277 | 0.7265 | 0.3935 |
| Res U net 4 attention with skip-connection in decoder, beta = 0 | 0.8273 | 0.8369 | 0.7041 | 0.7026 | 0.3895 |
| Res U net 4 attention, beta = 0.5 | 0.8618 | 0.8657 | 0.7572 | 0.7572 | 0.4307 |
| Res U net 4 attention with skip-connection in decoder, beta = 0.5 | 0.84014 | 0.8499 | 0.7230 | 0.7216 | 0.4053 |
| Res U net 4 attention, beta = 1 | 0.8636 | 0.8718 | 0.7593 | 0.7585 | 0.4168 |
| Res U net 4 attention with skip-connection in decoder, beta = 1 | 0.8522 | 0.8593 | 0.7420 | 0.7414 | 0.4170 |
| Res U net 4 attention, beta = 1.5 | 0.7786 | 0.7923 | 0.6321 | 0.6291 | 0.3297 |
| Res U net 4 attention with skip-connection in decoder, beta = 1.5 | 0.753 | 0.7685 | 0.5960 | 0.5921 | 0.4153 |
| Res U net 4 attention, beta = 2 | 0.87197 | 0.8788 | 0.7727 | 0.7722 | 0.4482 |
| Res U net 4 attention with skip-connection in decoder, beta = 2 | 0.861 | 0.8684 | 0.7556 | 0.7549 | 0.4538 |

Although there is no obvious trend in the validation accuracy, when it comes to testing on unseen case, the models with skip-connections in decoder consistently perform slightly worse than the ones containing skip-connections only in the encoder.

## 4.5.8. End of encoder for fine features input (proposed) or Decoder for fine features input (literature) inside of attention modules

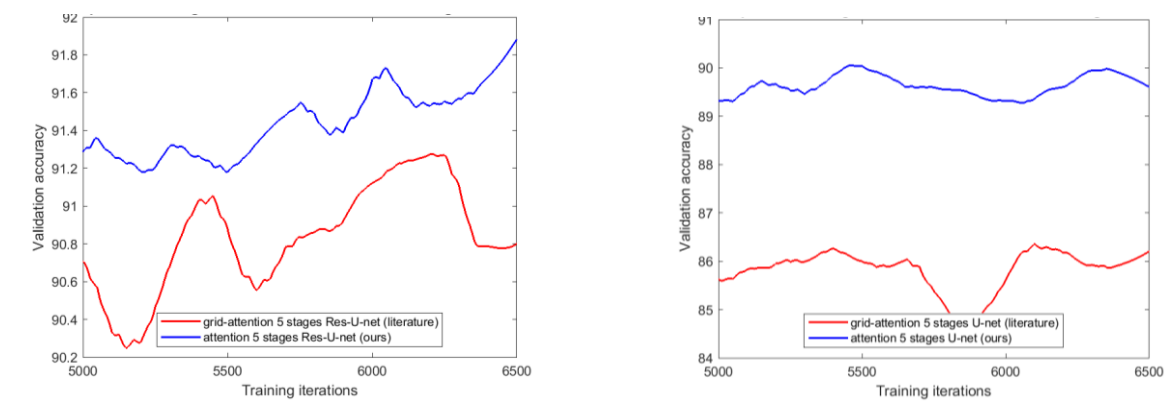

*Figure 28: Left: different attention connections in Res-U-net, asymmetric loss function, beta = 1; Right: different attention connection in U-net, asymmetric loss function, beta = 1*

*Table 8: Comparison of testing scores on unseen case between two different integration methods of attention modules*

| Model_names | Global_accuracy | Mean_accuracy | Mean_IoU | Weighted_IoU | Mean_BF_Score |
|---|---|---|---|---|---|
| Res-U-net 4 attention, beta = 1 (connections in literature) | 0.48723 (0.041398) | 0.51884 (0.036874) | 0.26432 (0.05546) | 0.25009 (0.058381) | 0.40885 (0.24161) |
| Res-U-net 4 attention, beta = 1 (proposed connections) | 0.86548 (0.021367) | 0.87167 (0.019422) | 0.763 (0.033812) | 0.76254 (0.03426) | 0.43682 (0.03707) |
| 5 U net 4 attention, beta = 1 (connections in literature) | 0.69517 (0.18311) | 0.70369 (0.16239) | 0.52924 (0.23524) | 0.5257 (0.2449) | 0.32201 (0.029123) |

| 5 U net 4 attention, beta = 1 (proposed connections) | 0.73711 (0.11586) | 0.74894 (0.10495) | 0.58074 (0.15471) | 0.57736 (0.15981) | 0.33891 (0.082998) |

It's very clearly here in either validation or testing, the attentional models using our proposed connection methods always achieve higher scores in all the evaluation metrices.

### 4.5.9. Element-addition attention and Vector-concatenation attention

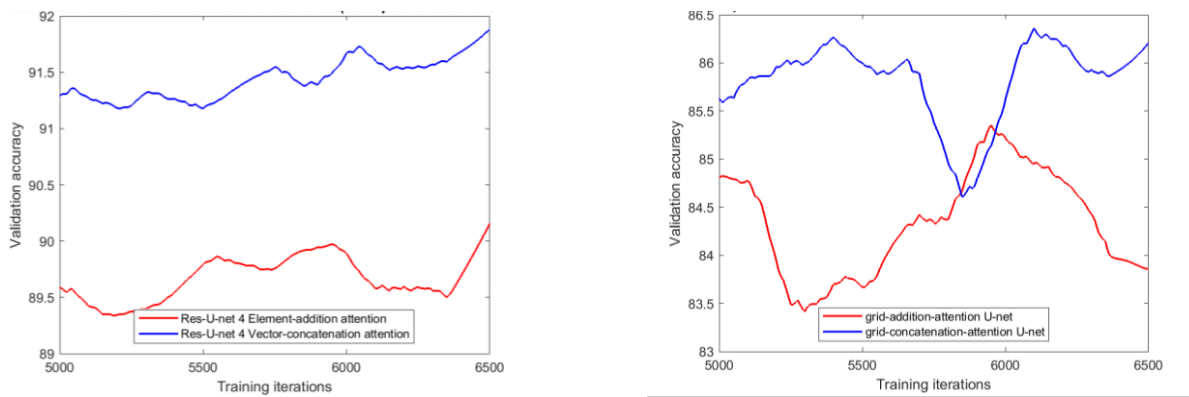

*Figure 29: Left: element-addition and vector-concatenation comparison in Res-U-net, asymmetric loss function, beta = 1; Right: element-addition and vector-concatenation comparison in plain U-net, asymmetric loss function, beta = 1*

*Table 9: Comparison of testing scores on unseen case between models using element-addition attention and vector-concatenation attention*

| Model_names | Global_accuracy | Mean_accuracy | Mean_IoU | Weighted_IoU | Mean_BF_Score |
|---|---|---|---|---|---|
| Res-U-net 4 element-addition attention, beta = 1 (proposed connection) | 0.78255 (0.07954) | 0.78953 (0.073199) | 0.64261 (0.11202) | 0.64136 (0.1147) | 0.30981 (0.079099) |
| Res-U-net 4 vector-concatenation attention, beta = | 0.86548 (0.021367) | 0.87167 (0.019422) | 0.763 (0.033812) | 0.76254 (0.03426) | 0.43682 (0.03707) |
| 1 (proposed connection) | | | | | |
| U-net 4 element-addition attention, beta = 1 (connection in literature) | 0.72865 (0.043621) | 0.72323 (0.043778) | 0.56675 (0.052545) | 0.56981 (0.05214) | 0.29439 (0.032983) |
| U-net 4 vector-concatenation attention, beta = 1 (connection in literature) | 0.69517 (0.18311) | 0.70369 (0.16239) | 0.52924 (0.23524) | 0.5257 (0.2449) | 0.32201 (0.029123) |

The trend in validation accuracy is very clear that the vector-concatenation attention modules bring better performances, however, combining with the testing results, it seems only in the models with short ship connections, the vector-concatenation attention performs better than the element-addition attention.

### 4.5.10. Cross-entropy loss function and Similarity measurement loss function

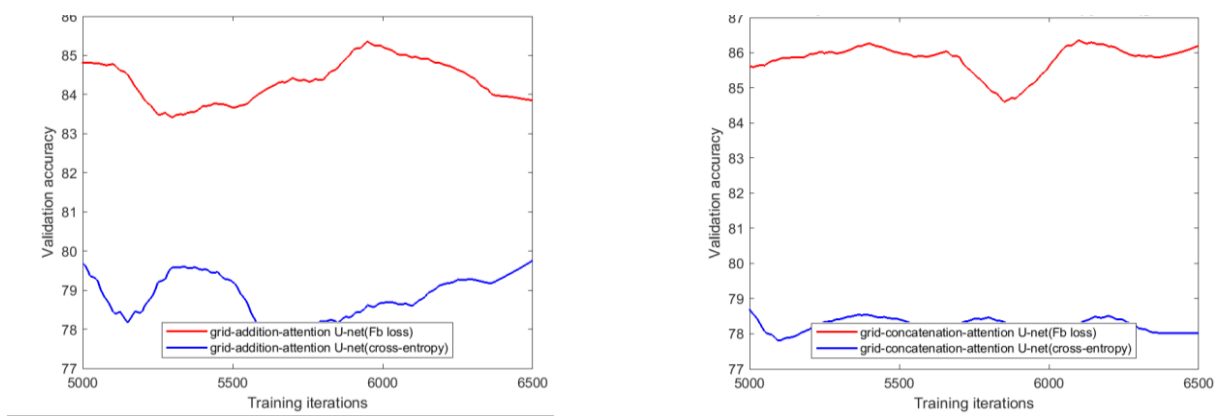

Figure 30: Left: cross entropy and asymmetric loss comparison in element-addition attention U-net; Right: cross-etnropy and asymmetric loss function comparison in vector-concatenation attention U-net

Table 10: Comparison of testing scores on unseen case between models using cross-entropy loss and asymmetric loss function

| Model_names | Global_accuracy | Mean_accuracy | Mean_IoU | Weighted_IoU | Mean_BF_Score |
|---|---|---|---|---|---|
| element-addition attention U-net, cross_entorpy (connection in literature) | 0.51316 (0.041369) | 0.50179 (0.0030964) | 0.25803 (0.02217) | 0.26488 (0.041878) | 0.35227 (0.27038) |
| element-addition attention U-net, beta = 1 (connection in literature) | 0.72865 (0.043621) | 0.72323 (0.043778) | 0.56675 (0.052545) | 0.56981 (0.05214) | 0.29439 (0.032983) |
| vector-concatenation attention U-net, cross_entorpy (connection in literature) | 0.52409 (0.10153) | 0.53846 (0.066618) | 0.29408 (0.10625) | 0.2875 (0.12268) | 0.55762 (0.24025) |
| vector-concatenation attention U-net, beta = 1 (connection in literature) | 0.69517 (0.18311) | 0.70369 (0.16239) | 0.52924 (0.23524) | 0.5257 (0.2449) | 0.32201 (0.029123) |

It's very clear in this series of experiments that the similarity measurement based loss function (asymmetric loss function) achieve much better performance than cross-entropy loss function due to its merit against the class imbalances, models using similarity loss function consistently achieve higher scores in both the validation and testing.

## 4.5.11. Spatial and mixed attention

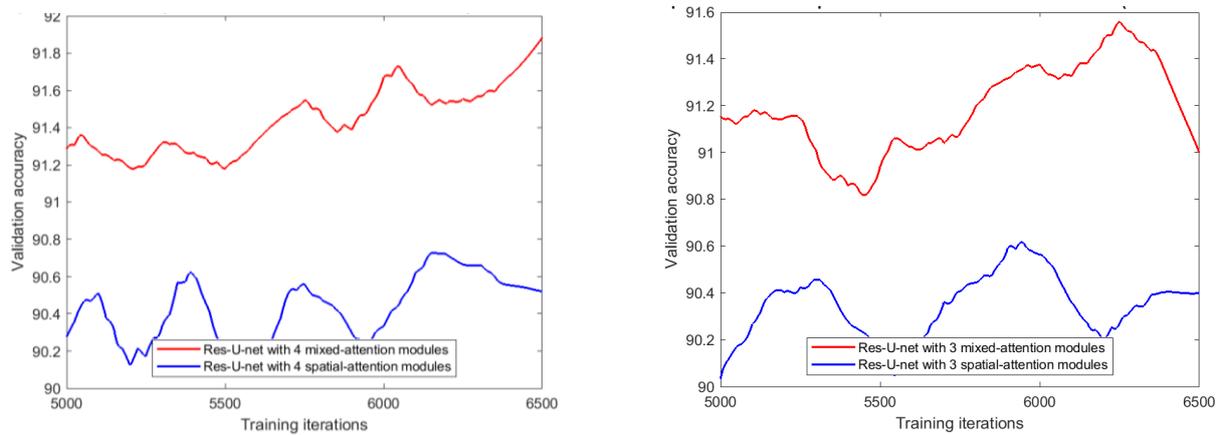

*Figure 31: Left: spatial attention and mixed attention comparison in Res-U-net with 4 attention modules; Right: spatial attention and mixed attention comparison in Res-U-net with 3 attention modules*

*Table 11: Comparison of testing scores on unseen case between models using spatial-attention and mixed-attention*

| Model_names | Global_accuracy | Mean_accuracy | Mean_IoU | Weighted_IoU | Mean_BF_Score |
|---|---|---|---|---|---|
| Res U net 3 spatial-attention, beta = 1 | 0.85422 (0.018758) | 0.85899 (0.020565) | 0.74574 (0.028299) | 0.74562 (0.027914) | 0.39741 (0.021745) |
| Res U net 3 mixed-attention, beta = 1 | 0.82784 (0.012076) | 0.836 (0.012276) | 0.70559 (0.017487) | 0.70454 (0.017502) | 0.39025 (0.028309) |
| Res U net 4 spatial-attention, beta = 1 | 0.85834 (0.018892) | 0.86242 (0.022268) | 0.75182 (0.029183) | 0.75187 (0.028454) | 0.39968 (0.050771) |
| Res U net 4 mixed-attention, beta = 1 | 0.86548 (0.021367) | 0.87167 (0.019422) | 0.763 (0.033812) | 0.76254 (0.03426) | 0.43682 (0.03707) |

Although the models with mixed-attention consistently achieve higher accuracy than the ones with spatial-attention, it seems in the testing, the mixed-attention is only better when there are 4 attention modules.

## 4.5.12. Effect of different hyperparameters of similarity measurement loss function (Fb beta loss function)

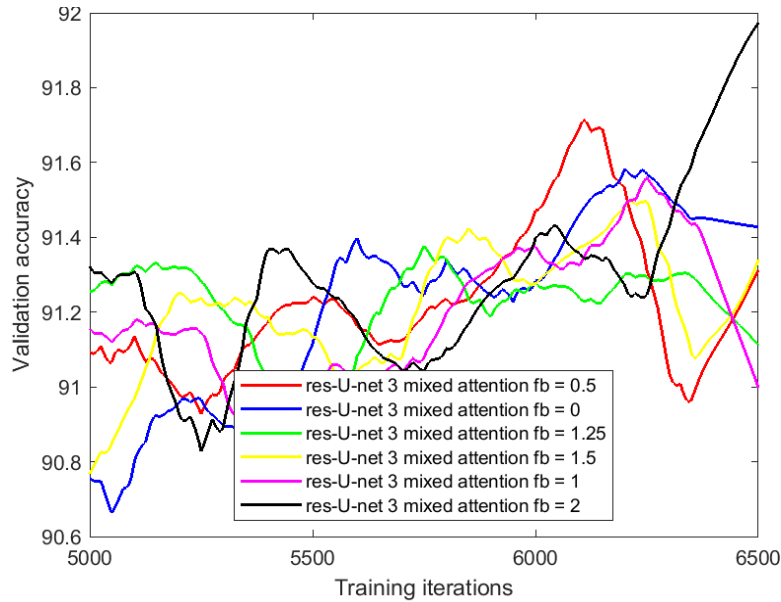

*Figure 32:beta values of asymmetric loss function comparison in Res-U-net with 3 mixed-concatenation attention modules*

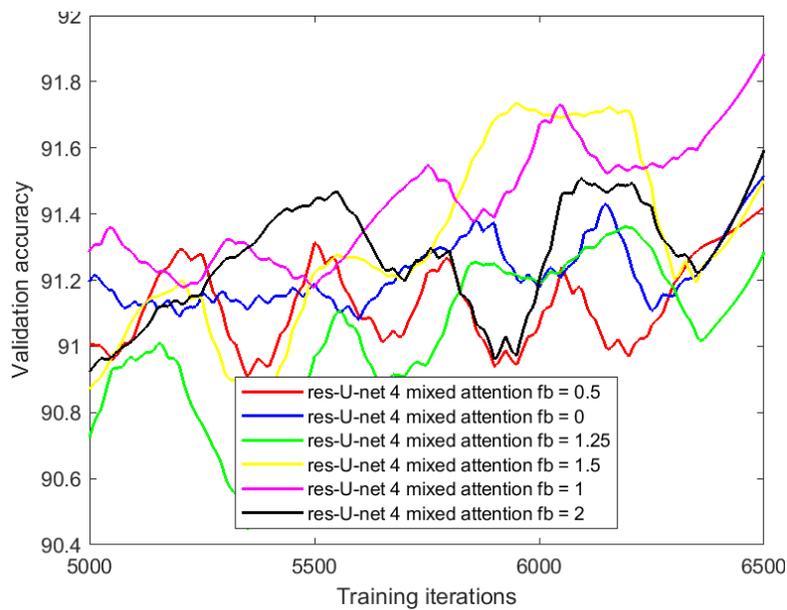

*Figure 33:beta values of asymmetric loss function comparison in Res-U-net with 4 mixed-concatenation attention modules*

*Table 12: Comparison of testing scores on unseen case among models using different beta values in the asymmetric loss function*

| Model_names | Global_accuracy | Mean_accuracy | Mean_IoU | Weighted_IoU | Mean_BF_Score |
|---|---|---|---|---|---|
| Res U net 3 mixed-attention, beta = 0 | 0.86231 (0.013754) | 0.86805 (0.011377) | 0.75776 (0.021753) | 0.75739 (0.022349) | 0.42871 (0.028282) |
| Res U net 4 mixed-attention, beta = 0 | 0.86712 (0.034489) | 0.87421 (0.032332) | 0.76594 (0.053605) | 0.76528 (0.054254) | 0.44481 (0.050977) |
| Res U net 3 mixed-attention, beta = 0.5 | 0.87816 (0.010588) | 0.88321 (0.0086827) | 0.78268 (0.016989) | 0.7825 (0.017431) | 0.46763 (0.020482) |
| Res U net 4 mixed-attention, beta = 0.5 | 0.85663 (0.024067) | 0.86461 (0.021937) | 0.74896 (0.037903) | 0.74807 (0.038471) | 0.4291 (0.043405) |
| Res U net 3 mixed-attention, beta = 1 | 0.82784 (0.012076) | 0.836 (0.012276) | 0.70559 (0.017487) | 0.70454 (0.017502) | 0.39025 (0.028309) |
| Res U net 4 mixed-attention, beta = 1 | 0.86548 (0.021367) | 0.87167 (0.019422) | 0.763 (0.033812) | 0.76254 (0.03426) | 0.43682 (0.03707) |
| Res U net 3 mixed-attention, beta = 1.25 | 0.85999 (0.028691) | 0.86732 (0.025829) | 0.75443 (0.045312) | 0.75367 (0.046018) | 0.42132 (0.035772) |
| Res U net 4 mixed-attention, beta = 1.25 | 0.85746 (0.017289) | 0.86483 (0.015288) | 0.75019 (0.027057) | 0.74946 (0.027578) | 0.42383 (0.006781) |
| Res U net 3 mixed-attention, beta = 1.5 | 0.81771 (0.038633) | 0.82872 (0.035067) | 0.68994 (0.058389) | 0.68799 (0.059596) | 0.37721 (0.046922) |
| Res U net 4 mixed-attention, beta = 1.5 | 0.82184 (0.064643) | 0.83162 (0.059886) | 0.69814 (0.09632) | 0.69644 (0.098116) | 0.41155 (0.052886) |

| Res U net 3 mixed-attention, beta = 2 | 0.84006 (0.035145) | 0.84882 (0.033425) | 0.72419 (0.053715) | 0.72302 (0.054315) | 0.39957 (0.051065) |
| Res U net 4 mixed-attention, beta = 2 | 0.8694 (0.010914) | 0.87542 (0.0091444) | 0.76886 (0.017361) | 0.76846 (0.017745) | 0.44649 (0.015911) |

Theoretically, higher beta values will bring more false positives and it seems the model with the highest beta value achieves the best validation accuracy in Res-U-net with 3 mixed-attention modules, however, in the Res-U-net with 4 mixed attention modules, the best performer is the model with the second highest beta value and the highest beta value is the second highest performer. The comparison in the testing is not clear either but it seems the biased loss function does change the ratio between the false positives and false negatives, for example, when there are 3 attention modules in Res-u-net, the testing accuracy reaches the highest when the beta value is 0.5; when there are 4 attention modules, the best performer is at the beta value 2.

## 4.5.13. Pre-activation mixed attention and After-activation mixed attention

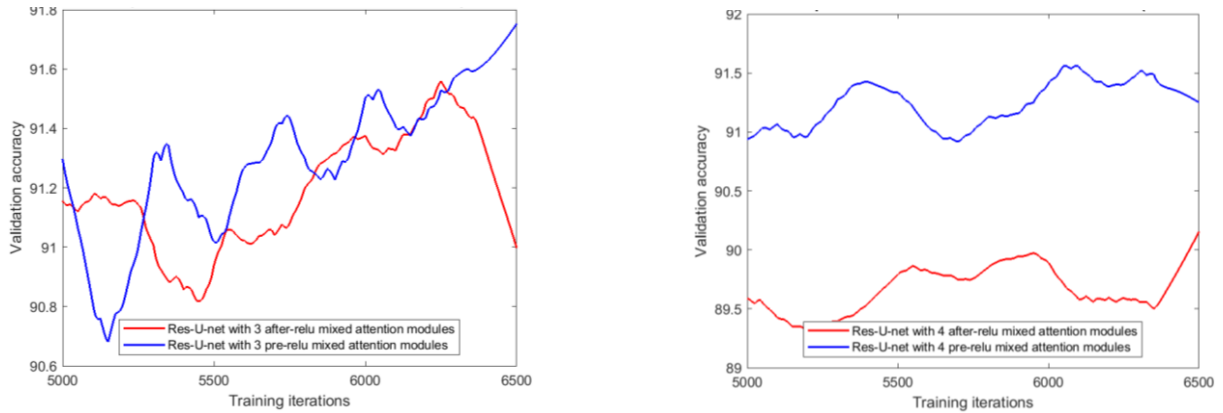

*Figure 34: Left: Pre-ReLU and After-ReLU comparison in Res-U-net with 3 attention modules; Right: Pre-ReLU and After-ReLU comparison in Res-U-net with 4 attention modules*

*Table 13: Comparison of testing scores on unseen case between models using pre-ReLU attention and after-ReLU attention*

| Model_names | Global_accuracy | Mean_accuracy | Mean_IoU | Weighted_IoU | Mean_BF_Score |
|---|---|---|---|---|---|
| Res U net 3 mixed after-relu attention, beta = 1 | 0.8508 (0.033025) | 0.85917 (0.029784) | 0.74015 (0.051675) | 0.73909 (0.052549) | 0.42102 (0.056542) |
| Res U net 3 mixed pre-relu attention, beta = 1 | 0.82784 (0.012076) | 0.836 (0.012276) | 0.70559 (0.017487) | 0.70454 (0.017502) | 0.39025 (0.028309) |
| Res U net 4 mixed after-relu attention, beta = 1 | 0.85057 (0.0056085) | 0.85861 (0.0059692) | 0.73937 (0.0083791) | 0.73847 (0.0083385) | 0.41545 (0.014895) |
| Res U net 4 mixed pre-relu attention, beta = 1 | 0.86548 (0.021367) | 0.87167 (0.019422) | 0.763 (0.033812) | 0.76254 (0.03426) | 0.43682 (0.03707) |

Although the pre-ReLU attention outperforms the after-ReLU attention in terms of validation accuracy, the Res-U-net with 3 pre-ReLU performs worse than the one with after-ReLU attention modules.

### 4.5.14. Comparison between proposed network and the state-of-the-art

This section shows statistical comparison of scores of our proposed network and the-state-of-art, due to space constraint, only the most representative networks are compared, across all the evaluation metrics, our model consistently outperforms the popular ones.

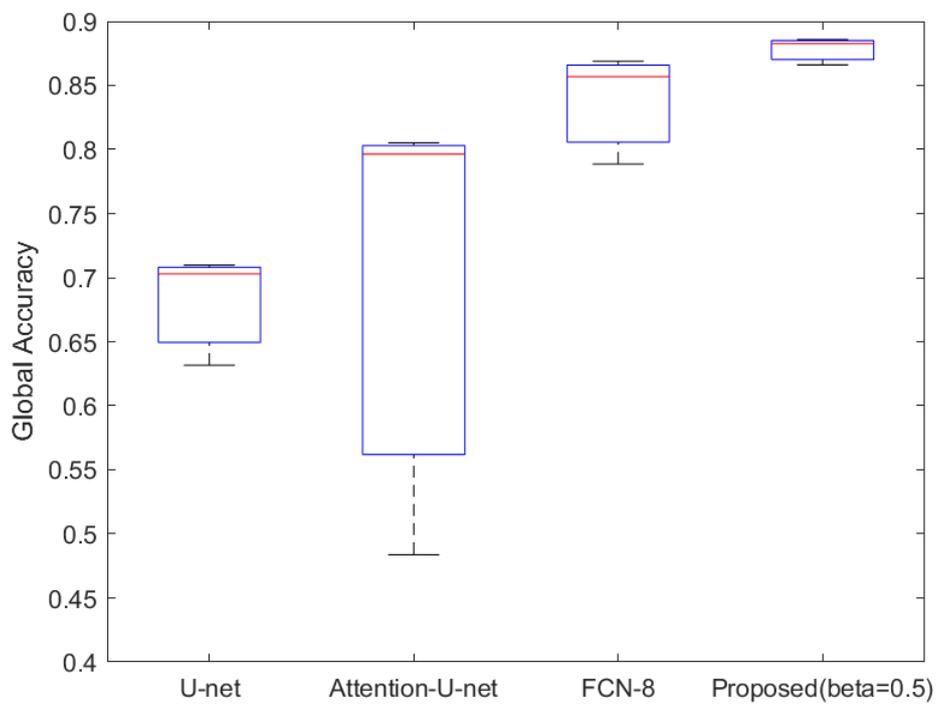

*Figure 35: Comparison between our proposed model and the state-of-the-art in terms of Global Accuracy*

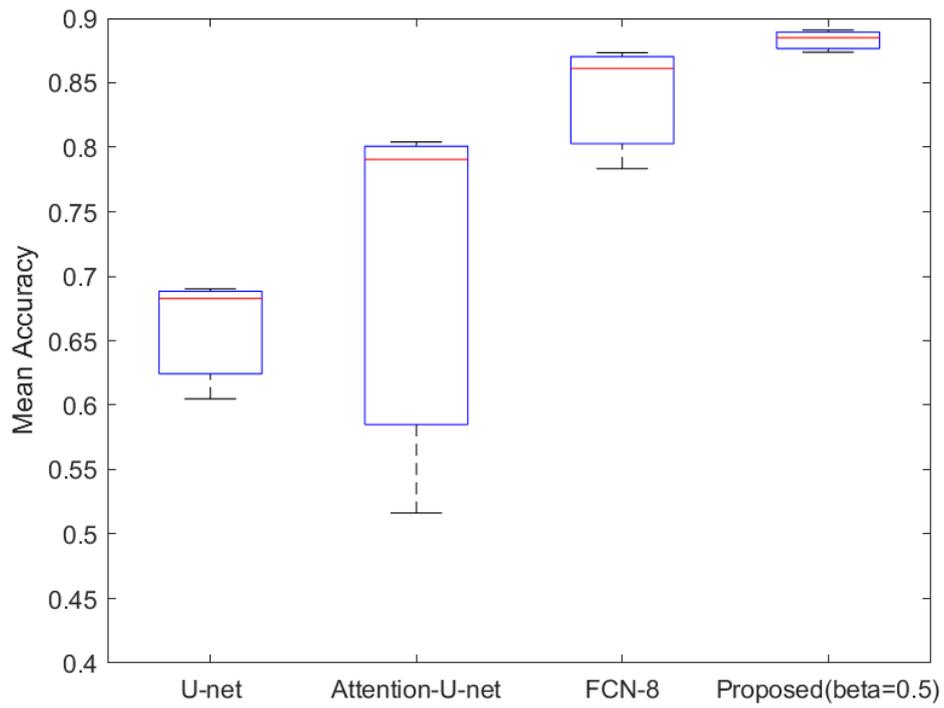

*Figure 36: Comparison between our proposed model and the state-of-the-art in terms of Mean Accuracy*

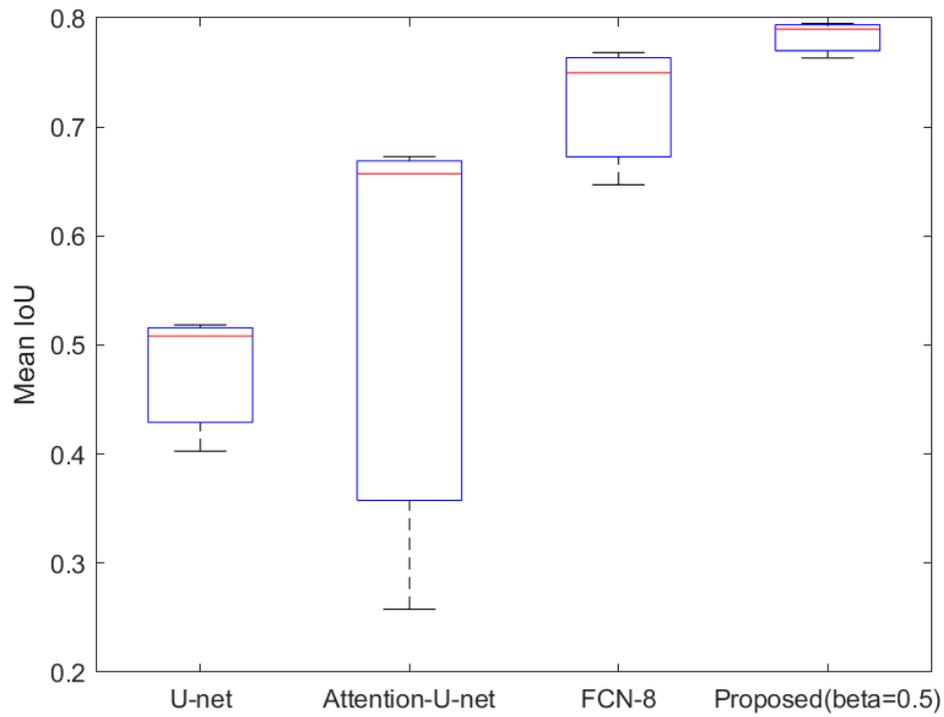

*Figure 37: Comparison between our proposed model and the state-of-the-art in terms of Mean IoU*

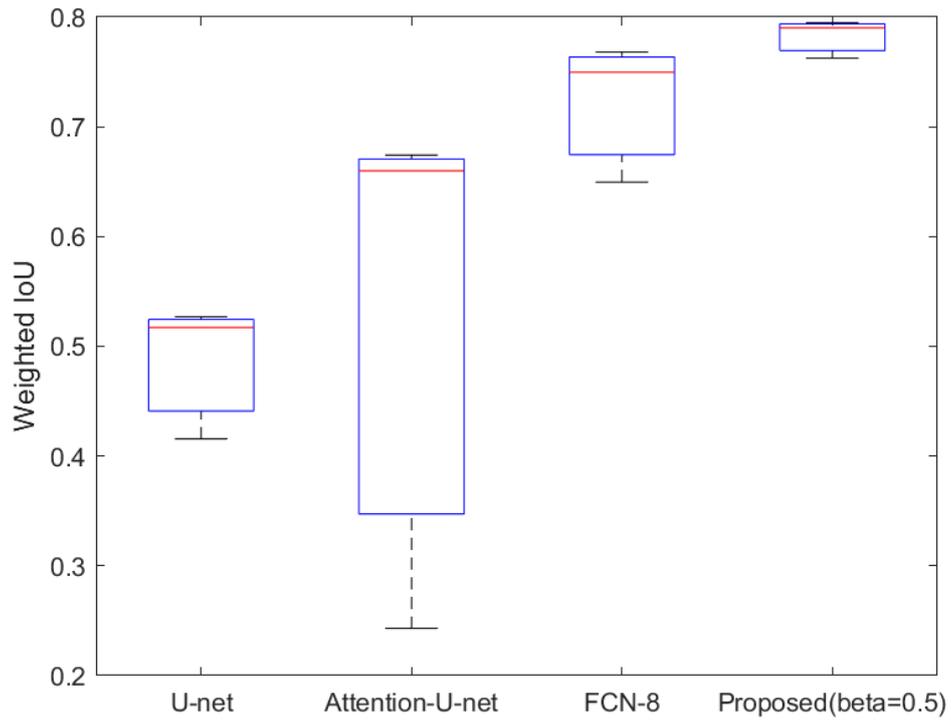

*Figure 38: Comparison between our proposed model and the state-of-the-art in terms of Weighted IoU*

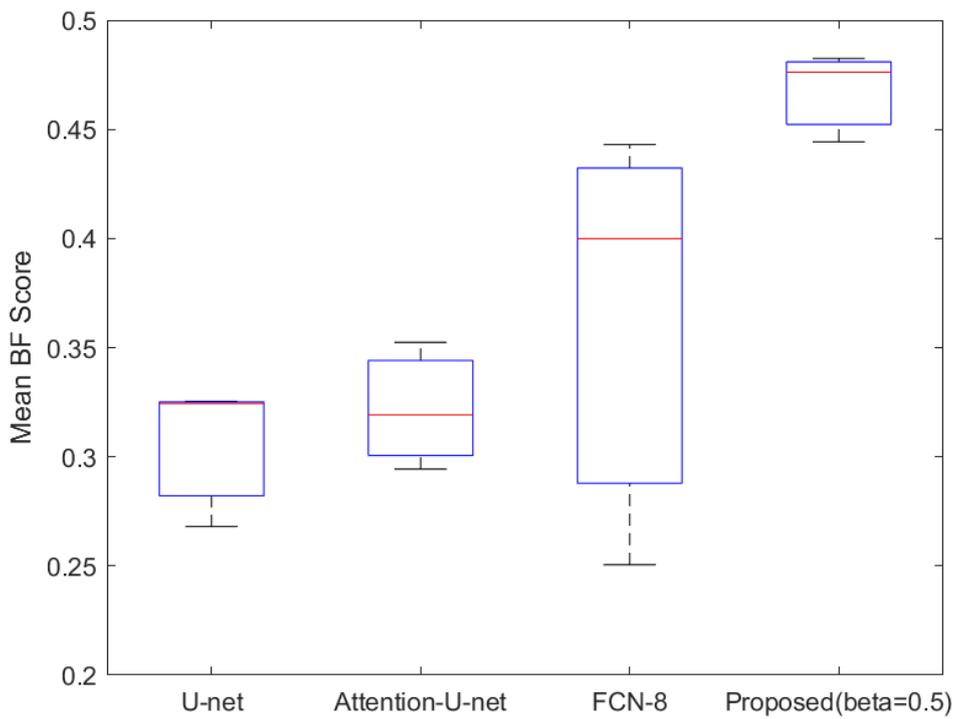

*Figure 39: Comparison between our proposed model and the state-of-the-art in terms of Mean BF score*

## 4.5.15. Visual illustration of trained attention mechanism

Here are visual results of the attention tested on 3 different inputs, as shown in figures below, with the depth goes deeper and deeper, the saliency areas concentrate and the surrounding areas get blurry, just like human visual systems doing focusing.

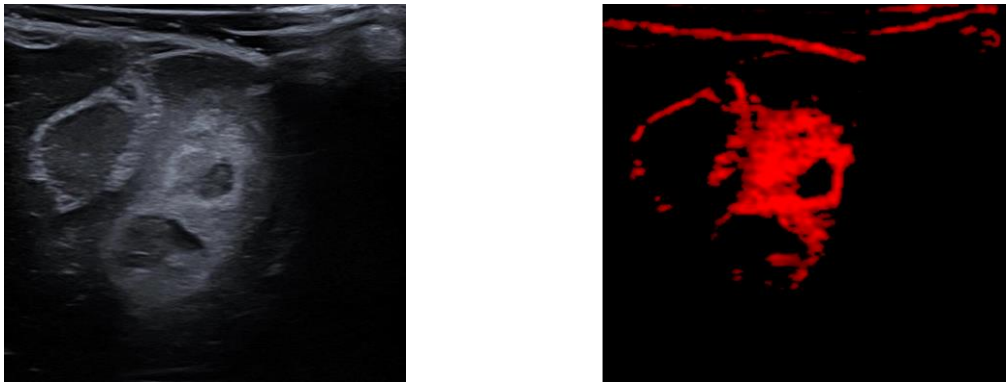

*Figure 40: Left: Original US image example 1, Right: attention in decoder stage 4 on example 1*

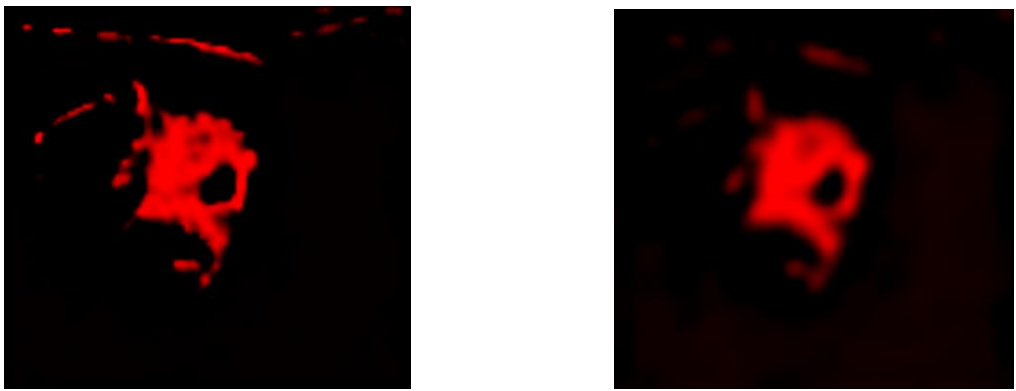

*Figure 41: Left: attention in decoder stage 3 on example 1, Right: attention in decoder stage 2 on example 1*

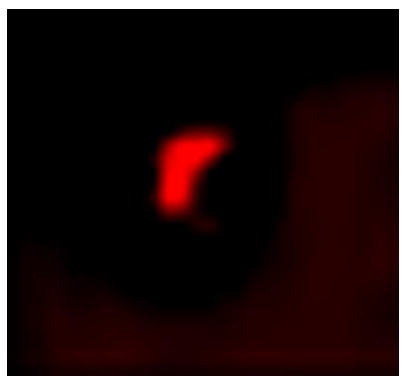

*Figure 42: attention in decoder stage 1 on example 1*

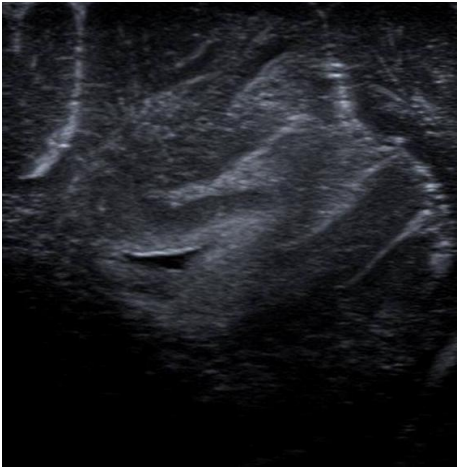 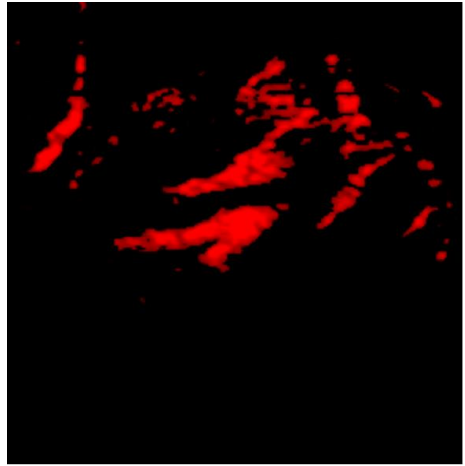

*Figure 43: Left: Original US image example 2, Right: attention in decoder stage 4 on example 2*

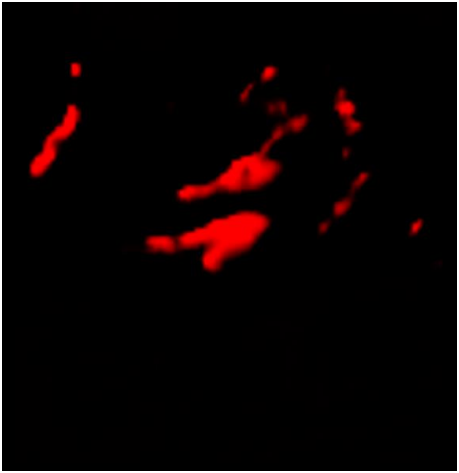 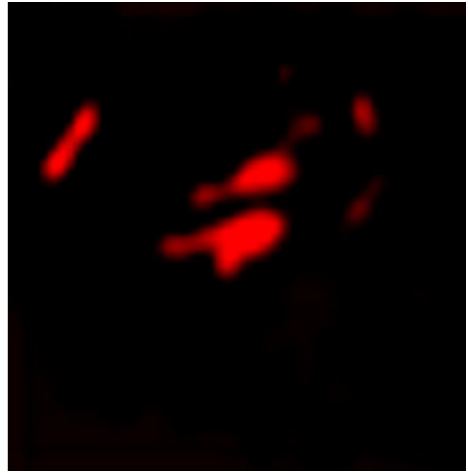

*Figure 44: Left: attention in decoder stage 3 on example 2, Right: attention in decoder stage 2 on example 2*

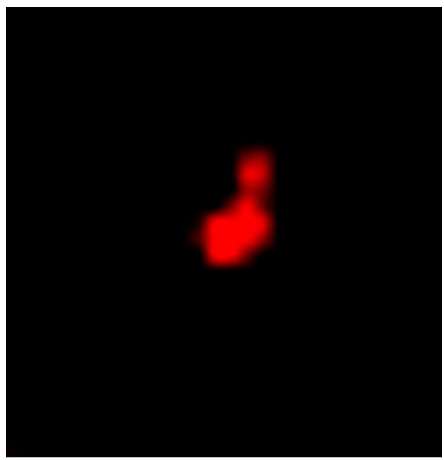

*Figure 45: attention in decoder stage 1 on example 2*

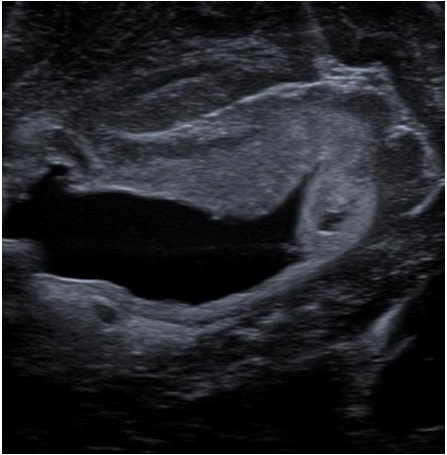 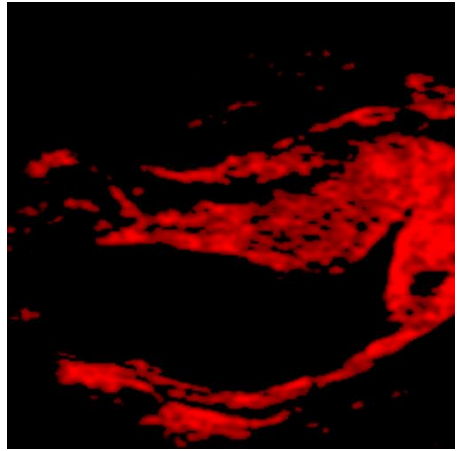

*Figure 46: Left: Original US image example 3, Right: attention in decoder stage 4 on example 3*

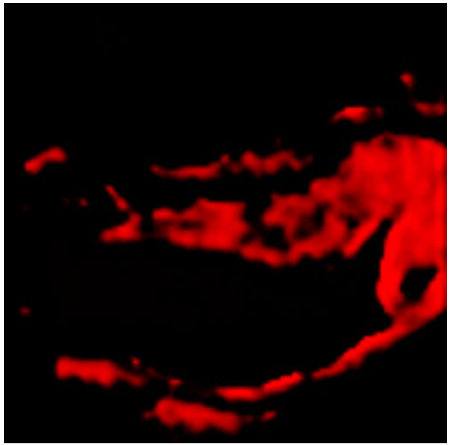 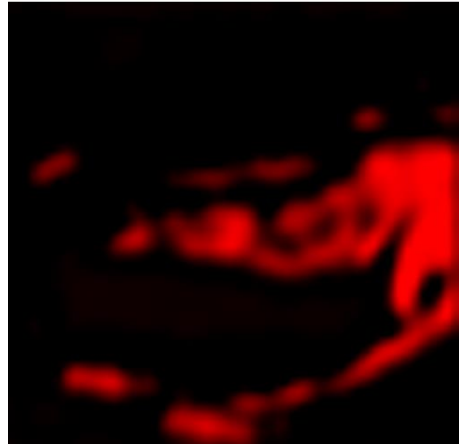

*Figure 47:Left: attention in decoder stage 3 on example 3, Right: attention in decoder stage 2 on example 3*

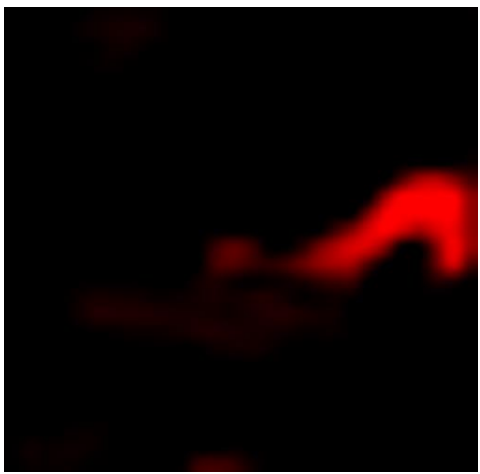

*Figure 48: attention in decoder stage 1 on example 3*

## 4.5.16. Visual comparison

### 4.5.16.1. Proposed architecture

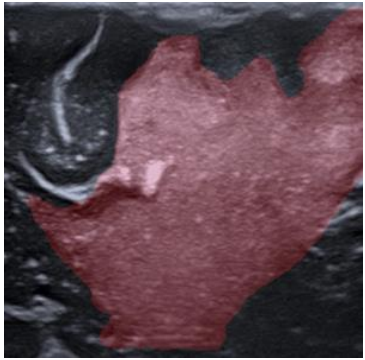 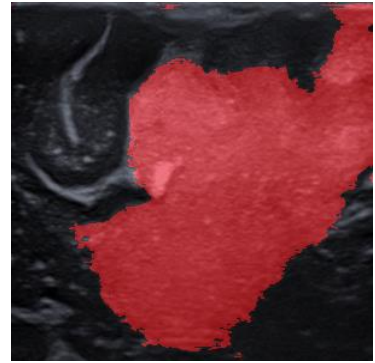

*Figure 49: Left: Ground truth of testing 1, Right: Segmentation on testing 1 using proposed architecture*

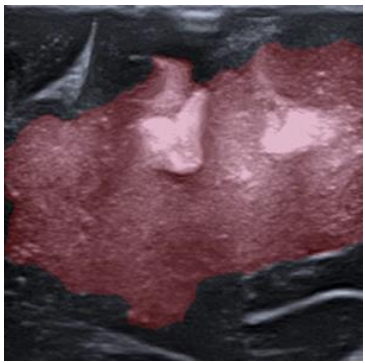 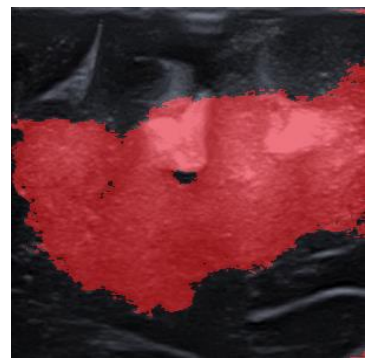

*Figure 50: Left: Ground truth of testing 2, Right: Segmentation on testing 2 using proposed architecture*

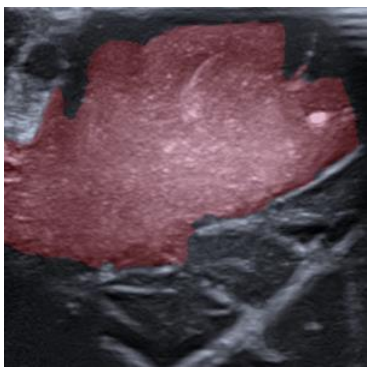 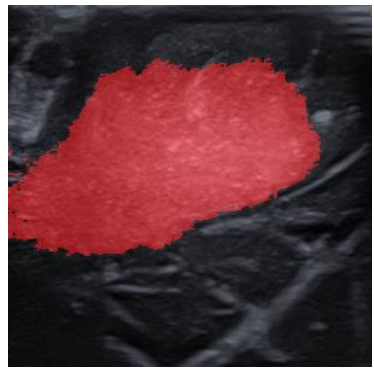

*Figure 51: Left: Ground truth of testing 3, Right: Segmentation on testing 3 using proposed architecture*

### 4.5.16.2. FCN-8s

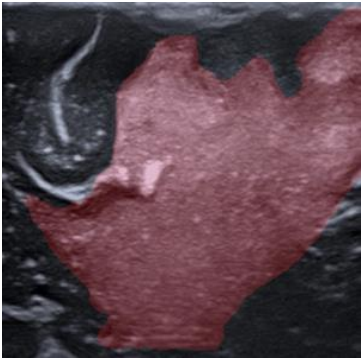 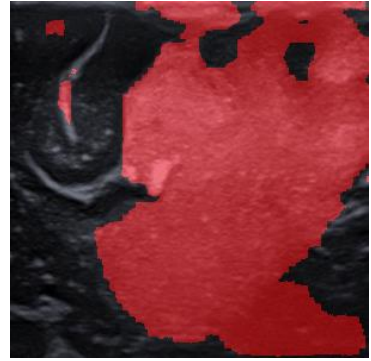

*Figure 52: Left: Ground truth of testing 1, Right: Segmentation on testing 1 using FCN-8s*

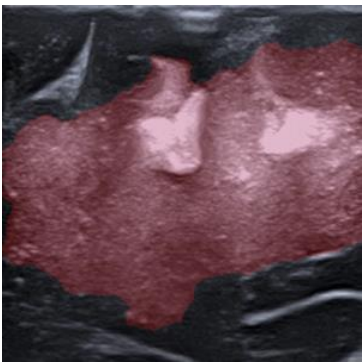 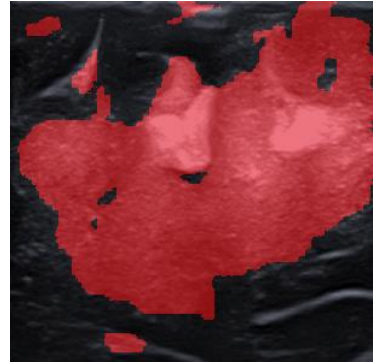

*Figure 53: Left: Ground truth of testing 2, Right: Segmentation on testing 2 using FCN-8s*

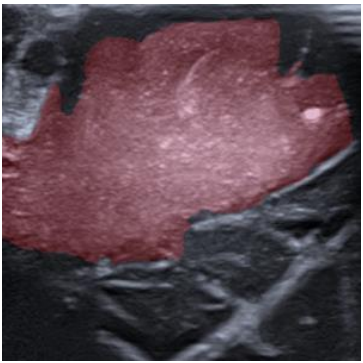 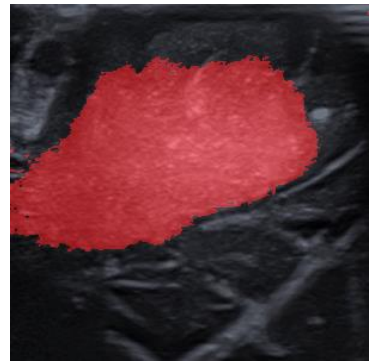

*Figure 54: Left: Ground truth of testing 3, Right: Segmentation on testing 3 using FCN-8s*

### 4.5.16.3. U-net

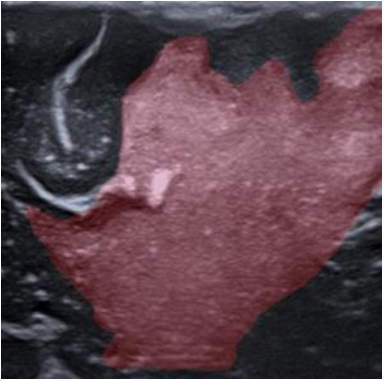 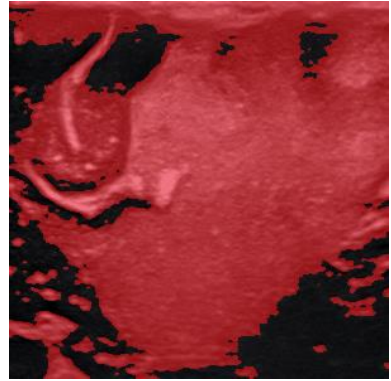

*Figure 55: Left: Ground truth of testing 1, Right: Segmentation on testing 1 using U-net*

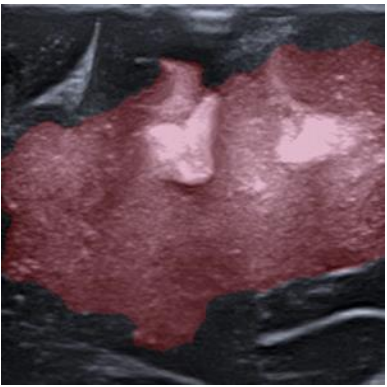 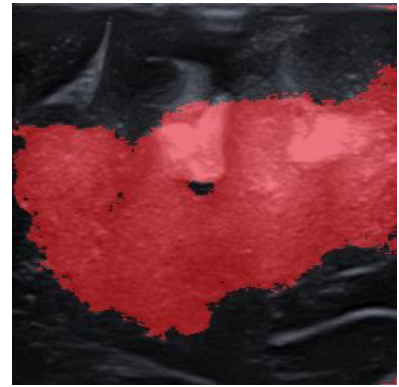

*Figure 56: Left: Ground truth of testing 2, Right: Segmentation on testing 2 using U-net*

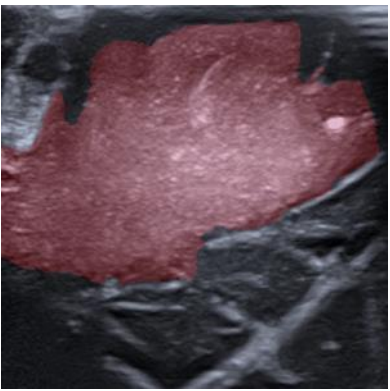 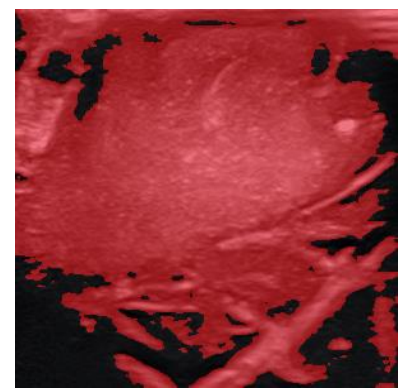

*Figure 57: Left: Ground truth of testing 3, Right: Segmentation on testing 3 using U-net*

# Chapter 5   Conclusion

**5.1. Conclusion**

Two different approaches were explored in the present project to design a novel CAD system to automatically diagnose GB tumours in intraoperative US images: classification based on patches and classification at pixel-level directly. In the first patch-based approach, the state-of-the-art deep neural networks were fine-tuned on our multi-scale multi-direction patches, a parallel multi-scale distinct blocks classification was also proposed to speed up classification process. In the second approach, a novel neural network was proposed and it outperforms the state-of-the-art semantic segmentation methods on our validation and unseen testing data, this newly proposed neural network is referred as "Attention-Res-U-net with asymmetric loss function". With our newly designed vector-concatenation-mixed attention and our new integration method of attention modules in long-skip connections, the system can eliminate the false positives significantly comparing to the previous semantic segmentation frameworks as show in **4.5.16.**

**5.2.   Future work**

The Attention-Res-U-net with asymmetric loss function can be extended into a more attentional version by adding attention modules in the short skip-connections in the encoder as well, however, due to time constraint, this configuration is not completed and tested. Besides, more investigations into the asymmetric loss function can be done to understand better the effect of beta values. Later with more data collected, more classes can also be introduced.

Eventually, this attention mechanism can be used to design an unsupervised active learning framework since the attention can detect the saliency automatically and this characteristic can be used to automatically label the training data.

2018.

[62] O. Oktay *et al.*, "Attention U-Net: Learning Where to Look for the Pancreas," no. 2015, p. 2017, 2018.

[63] S. O. Oprea, "A Review on Deep Learning Techniques Applied to Semantic Segmentation," pp. 1–23.

[64] G. Csurka, D. Larlus, and F. Perronnin, "What is a good evaluation measure for semantic segmentation?," *Procedings Br. Mach. Vis. Conf. 2013*, p. 32.1-32.11, 2013.

[65] M. Rajchl, N. Pawlowski, D. Rueckert, P. M. Matthews, and B. Glocker, "NeuroNet: Fast and Robust Reproduction of Multiple Brain Image Segmentation Pipelines," no. Midl, pp. 1–9, 2018.
94